\documentclass[twocolumn,showpacs,prb,citeautoscript,amsmath,amssymb,floatfix]{revtex4-2}
\usepackage{amsmath,amssymb,color}
\usepackage{bm}
\usepackage{graphicx}
\usepackage{psfrag}
\usepackage{bbold}
\usepackage{bbm}
\newcommand{\bd}{\bm}

\begin{document}

\title{Fermion condensation  in a generalized  Hatsugai-Kohmoto model with momentum-mixing Landau interactions}

\author{Jan Heinrich, Andreas R\"{u}ckriegel, and  Peter Kopietz}
  
\affiliation{Institut f\"{u}r Theoretische Physik, Universit\"{a}t
  Frankfurt,  Max-von-Laue Stra{\ss}e 1, 60438 Frankfurt, Germany}

\date{May 20, 2026}

 \begin{abstract}
The Hatsugai-Kohmoto (HK) model is an exactly solvable electronic lattice model where the interaction
between electrons with opposite spin is diagonal in momentum space. 
We generalize the HK model by introducing momentum-mixing Landau interactions. 
Within a  self-consistent mean-field analysis we find that the ground state of this model
exhibits a partially flat energy band, in agreement with the 
fermion condensation scenario proposed by
Khodel and Shaginyan [JETP Lett. {\bf{51}}, 553 (1990)].
Inspired by  Andersons pseudospin formulation of BCS theory,
we show that
the  HK model with Landau interactions can be mapped onto 
a generalized Ising model where each site of the reciprocal lattice hosts two Ising spins.  In the pseudospin picture the emergence of a partially flat electronic band corresponds to the smoothing of a
magnetic domain wall. 
Moreover, 
guided by the pseudospin picture, we propose an exactly solvable variant of the HK model which
has a unique  ground state for all densities.

\end{abstract}


\maketitle


\section{Introduction}

In 1992 Hatsugai and Kohmoto (HK) introduced an exactly solvable  toy model for correlated electrons 
on a lattice which exhibits a Mott transition \cite{Hatsugai92}. The HK model is defined via the following second quantized Hamiltonian,
 \begin{equation}
 {\cal{H}}_{\rm HK} = 
 \sum_{\bd{k}}  \left[ \epsilon_{\bd{k}} ( n_{ \bd{k} \uparrow} + n_{\bd{k} \downarrow} )
 + U  n_{\bd{k} \uparrow} n_{\bd{k} \downarrow} \right],
 \label{eq:HK}
 \end{equation}
where the momentum sum is over the first Brillouin zone of a $d$-dimensional Bravais lattice,
$\epsilon_{\bd{k}}$ is some energy dispersion, and
the operator $n_{\bd{k} \sigma} = c^{\dagger}_{\bd{k} \sigma} c_{\bd{k} \sigma}$ 
counts the number of electrons with spin $\sigma = \uparrow , \downarrow$
occupying the reciprocal lattice site with crystal momentum $\bd{k}$. Here $c_{\bd{k} \sigma}$ and $c^{\dagger}_{\bd{k} \sigma} $ are the usual 
annihilation and creation operators of electrons.
The HK model (\ref{eq:HK}) describes the competition between the kinetic energy
encoded in the band dispersion $\epsilon_{\bd{k}}$, and the energy cost $U$ for occupying 
the reciprocal lattice site $\bd{k}$ by two electrons with opposite spin. Because
the Hamiltonian (\ref{eq:HK}) is block diagonal in momentum space and diagonal in the occupation number basis, all eigenstates and eigenenergies can be immediately written down. Nevertheless, the phase diagram of the HK model~(\ref{eq:HK}) 
is rather non-trivial and 
exhibits a Mott transition and several metallic phases, see Sec.~\ref{sec:phasesHK}.
Due to the complete absence of momentum mixing the HK model does not describe realistic electron-electron interactions in solids so that
for many years the HK model
has  recieved only limited attention as a toy model exhibiting a Mott transition \cite{Continentino94,Vitoriano00,Yeo19,Phillips20} with some  unphysical 
properties~\cite{Guerci25}.
However, recently  various modifications of the HK model including momentum mixing have been proposed which reproduce experimentally observed features of strongly correlated electrons~\cite{Li22,Zhao23,Wang24,Worm24,Shi25,Tenkila25,Mai26,
Manning26}.
Surprisingly, rather simple forms of momentum mixing involving the coupling of only a  few special momenta
are sufficient to reproduce hallmarks of strong correlation physics such as partially flat bands or non-Fermi liquid physics.

In this work we consider a modification of the HK model including momentum mixing   of a macroscopic number of states.  Our model can be obtained from the HK Hamiltonian (\ref{eq:HK}) by adding   a Landau type of interaction,
 \begin{align}
 {\cal{H}}_{\rm HKL} & = {\cal{H}}_{\rm HK}
 +
 \frac{1}{2 \cal{V}} 
 \sum_{\bd{k} \bd{k}^{\prime} \sigma \sigma^{\prime} } f_{\bd{k}  \bd{k}^\prime }  n_{\bd{k} \sigma} n_{\bd{k}^{\prime} \sigma^{\prime}},
 \label{eq:HKL}
 \end{align}
where ${\cal{V}}$ is the volume of the system and
the interaction parameters $f_{\bd{k} \bd{k}^\prime}$ couple occupation numbers with momenta $\bd{k}$ and $\bd{k}^\prime$. 
We refer to the model defined in Eq.~(\ref{eq:HKL}) as the
Hatsugai-Kohmoto-Landau (HKL) model.
A similar model has been studied previously in Ref.~[\onlinecite{Lidsky98}]; however these authors considered only a special type of 
Landau interaction  $f_{\bd{k}  \bd{k}^\prime }$ proportional to 
$\delta ( | \bd{k} | - | \bd{k}^{\prime}  | )  $. Here we study more general interactions which are finite as long as $| \bd{k} - \bd{k}^{\prime} | \lesssim 1/ \lambda $,
where the length scale $\lambda$ is large compared with the lattice spacing.
Because the Landau interaction involves only the occupation numbers in  momentum space all 
parts of the Hamiltonian (\ref{eq:HKL}) still commute.
Of course, more realistic interactions do not have this property. Nevertheless, it is interesting to see how the momentum mixing modifies the phase diagram of the HK model. Because the HKL model defined in Eq.~(\ref{eq:HKL}) cannot be solved exactly,
we will determine its phase diagram within the self-consistent
mean-field approximation. 

The rest of this work is organized as follows: To define our notation and set the stage for the rest of this work, we present in Sec.~\ref{sec:phasesHK} a systematic derivation of the ground state phase diagram of the HK model for fixed density as well as for fixed chemical potential. In Sec.~\ref{sec:pseudospin} we show that, after shifting the total energy and the chemical potential, the HK model can be expressed  in terms of suitably defined pseudospins. 
The pseudospin picture is useful to derive a modified HK model which does  not exhibit any ground state degeneracies. Moreover, in terms of pseudospins the HKL model (\ref{eq:HKL}) 
becomes a generalized Ising model where two Ising spins are attached to each site of the reciprocal lattice. In Sec.~\ref{sec:fermicond} we present a self-consistent mean-field analysis of the
pseudospin-HKL model and show that the Landau interaction 
destroys the discontinuites in the momentum distribution and generates partially flat bands. 
Finally, in Sec.~\ref{sec:summary} we summarize our results and present our conclusions.
In the Appendix we derive the momentum distribution of the HK model and the magnetic equation of state of the corresponding pseudospin HK model.

\section{Phases of the repulsive HK model at zero temperature}
\label{sec:phasesHK}
 
To set the stage for the calculations in the rest of this work, we review in this section the phase diagram of the 
HK model at zero temperature as a function of
density $\rho$ and
interaction $U$. We also derive the phase diagram as a function of 
chemical potential $\mu$ and bandwidth in units of $U$,
where the phase boundaries  are independent of the dimensionality of the system.

\subsection{Classification of ground states}

The HK Hamiltonian (\ref{eq:HK})  can be written as a direct sum ${\cal{H}}_{\rm HK}
= \sum_{\bd{k}} {\cal{H}}_{\bd{k}}$, where
 \begin{equation}
 {\cal{H}}_{\bd{k}} = \epsilon_{\bd{k}} n_{\bd{k}} 
 + U n_{\bd{k} \uparrow} n_{\bd{k} \downarrow} 
 \end{equation}
acts on the Fock space associated with a given reciprocal lattice site $\bd{k}$, and
 \begin{equation}
  n_{\bd{k}} = n_{\bd{k} \uparrow} + n_{\bd{k} \downarrow}
  \end{equation}
represents the total occupation number of this site. 
For simplicity we assume that $\epsilon_{\bd{k}}$ is symmetric 
around zero energy so that the corresponding density of states $\nu ( \epsilon )$ satisfies
$ \nu (  \epsilon ) = \nu ( - \epsilon )$.
The eigenstates of ${\cal{H}}_{\bd{k}}$ are given by the eigenstates of $n_{\bd{k} \sigma}$ which are given by
 \begin{subequations}
 \label{eq:basisoccupation}
 \begin{align}
 | 0 \rangle & ,
 \\
 |  1 \rangle_{\bd{k} \uparrow} & = c^{\dagger}_{\bd{k} \uparrow} | 0 \rangle,
 \\
 |   1 \rangle_{\bd{k} \downarrow} & = c^{\dagger}_{\bd{k} \downarrow} | 0 \rangle,
 \\
 | 2 \rangle_{\bd{k}} & =   c^{\dagger}_{\bd{k} \uparrow} c^{\dagger}_{\bd{k} \downarrow} | 0 \rangle,
 \end{align}
 \end{subequations}
where the vacuum state $| 0 \rangle$ satisfies
$c_{\bd{k} \sigma} | 0 \rangle =0$ for all momenta $\bd{k}$ and spin projections  $\sigma = \uparrow, \downarrow$. The  operator $n_{\bd{k}}$ has eigenvalues $0$, $1$, and $2$
so that in the  basis (\ref{eq:basisoccupation}) the block
${\cal{H}}_{\bd{k}}$ is represented by the diagonal matrix
 \begin{equation}
 {\mathbf{H}}_{\bd{k}} =
  \left( \begin{array}{cccc} 0  & 0 & 0 & 0 \\
  0& \epsilon_{\bd{k}} & 0 &  0\\
  0 & 0 & \epsilon_{\bd{k}}  & 0 \\
  0 &  0 & 0 &  2 \epsilon_{\bd{k}}  + U \end{array}
  \right).
  \label{eq:HKmatrix1}
  \end{equation}  
The set of all ground state phases of
the HK model can be labelled by the power of the set $ \{ 0,1,2 \}$ of eigenvalues of
$n_{\bd{k}}$  (i.e. by the set of all subsets) without the empty set.
Associating  the eigenvalue $0$ with the label H (hole = empty state), the eigenvalue $1$ with  S (singly occupied state) and the
eigenvalue $2$ with $D$ (doubly occupied state), the possible phases of the HK model 
can be labelled by the following seven subsets of the set $\{ H,S,D \}$,
 \begin{equation}
  \{H \}  , \{S \} , \{D \} , \{ H, S \}, \{ H, D \}, \{ S,D \}, \{H,S,D \} .
 \end{equation}
Here $\{ H \}$ represents an empty band without particles corresponding to  $n_{\bd{k}} =0$ for all
$\bd{k}$. Similarly, the set $\{ D \}$ represents a completely filled band where all states are doubly occupied and hence $n_{\bd{k}} =2$ for all $\bd{k}$. 
The set  $\{H,D \}$ represents a phase with only holes and doubly occupied states, which is the ground state of the system for $U \leq  0$. Here we focus on the case of repulsive interaction $U > 0$. Depending on the density and the value of the interaction, the system can then be found in one of the four phases associated with the four subsets 
$\{H,S,D \}, \{ H,S \}, \{ S,D \}, \{ S \}$. Let us discuss the phases in more detail:


\subsubsection{Phase HSD: holes, singly and doubly occupied states}
\label{subsubsec:ESD}

\begin{figure}[tb]
 \begin{center}
  \centering
 \includegraphics[width=0.45\textwidth]{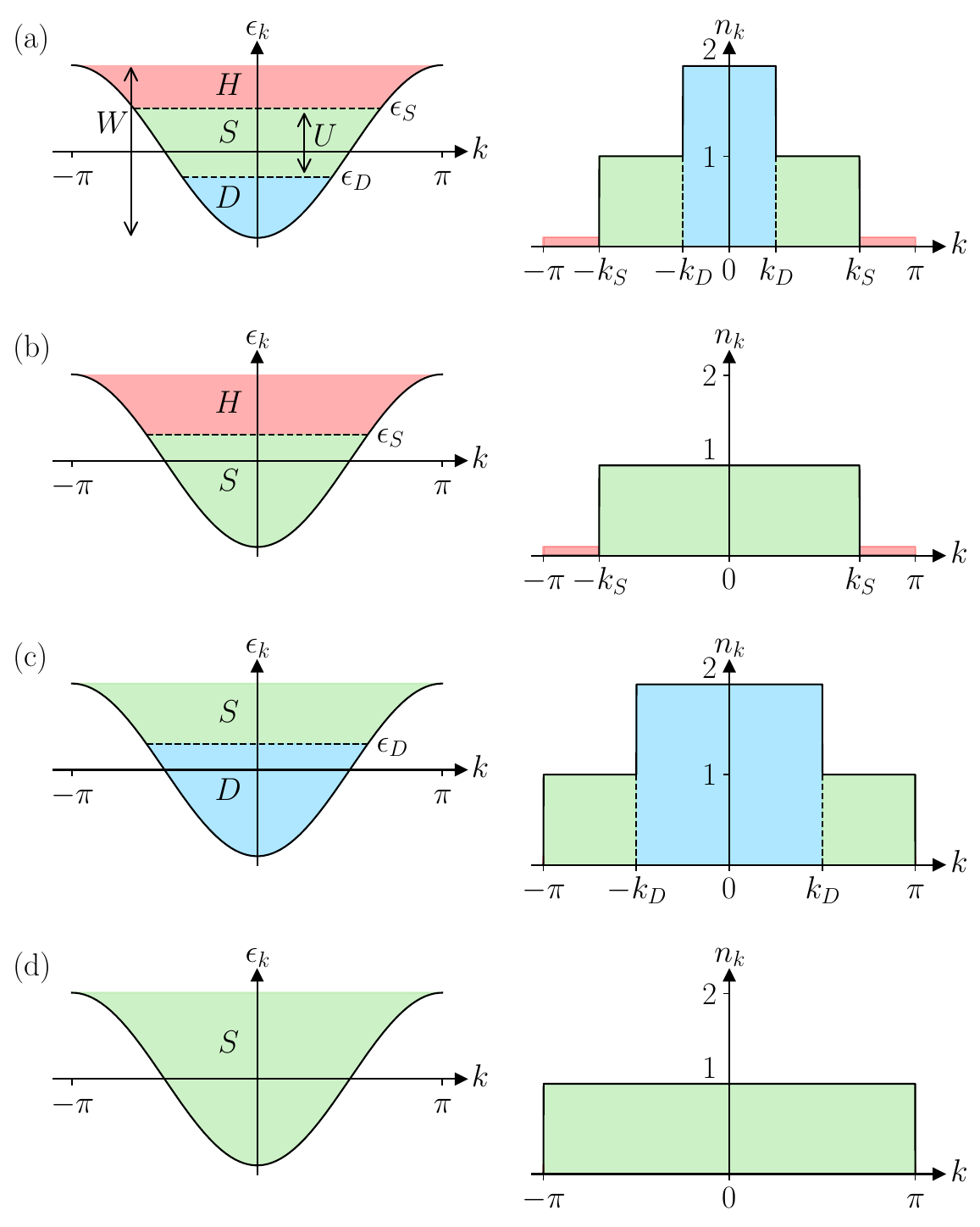}
   \end{center}
  \vspace{-4mm}
  \caption{%
(a) Left figure: ground state occupation of a one-dimensional energy band in the phase HSD where all three types of occupations coexist in the band. Here and in all subsequent figures empty states (holes) are shaded red, singly occupied states are shaded green,  and doubly occupied states are shaded blue.  
The horizontal dashed lines mark the Fermi energies $\epsilon_D$ and $\epsilon_S$ where the occupation changes discontinuously. 
Right figure: momentum distribution in the HSD phase. The 
vertical dashed lines mark the positions of the Fermi momenta $k_D$ and $k_S$.
All wavevectors are measured in units of the inverse lattice spacing.
(b)-(d) are analogous illustrations of the phases HS, SD, and S. 
}
\label{fig:phaseHSD}
\end{figure}

In this phase the ground state is an antisymmetrized product of all three types of momentum eigenstates corresponding to empty, singly,  and doubly occupied states.
Explicitly, the ground state in the phase HSD is given by
 \begin{equation}
 | E^{\rm{HSD}}_0 , \{ \sigma_{\bd{k}} \} \rangle =
  \left(  \prod_{ \epsilon_D < \epsilon_{\bd{k}} < \epsilon_S }c^{\dagger}_{\bd{k}
    \sigma_{\bd{k}} } \right) 
    \left(  \prod_{  \epsilon_{\bd{k}} < \epsilon_D } c^{\dagger}_{\bd{k}
    \uparrow } c^{\dagger}_{\bd{k} \downarrow } \right)
   | 0 \rangle,
   \label{eq:EHSD}
  \end{equation}
where $\epsilon_D$ is the energy of the most energetic doubly occupied  momentum state
while
 \begin{equation}
\epsilon_S = \epsilon_D + U
 \end{equation}
is the energy of the most energetic singly occupied momentum state.  
The spins $\sigma_{\bd{k}}$ of the electrons associated with
singly occupied states are arbitrary, so that the ground state 
(\ref{eq:EHSD}) is degenerate.  
A schematic picture  of the ground state configuration and the associated momentum distribution in one dimension is shown in Fig.~\ref{fig:phaseHSD}~(a).
This phase is characterized by two Fermi surfaces: an inner Fermi surface $\bd{k}_D$ defined by
$\epsilon_{\bd{k}_D} = \epsilon_D$ 
separating doubly occupied from singly occupied states, and an outer Fermi surface $\bd{k}_S$
defined by $\epsilon_{\bd{k}_S} = \epsilon_S$
separating singly occupied from empty states. The fact that the difference $ \epsilon_S - \epsilon_D$ of Fermi energies $\epsilon_S$ and $\epsilon_D$  is exactly given by the interaction $U$ can be seen as follows:
Suppose $ \epsilon_S - \epsilon_D < U$;  then the energy of the system can be lowered by
moving one of the electrons from a most energetic doubly occupied state to a previously empty state, because  the additional kinetic energy $\epsilon_S - \epsilon_D$ is smaller than the interaction  penalty $U$. The  original state therefore cannot be the ground state of the system. 
On the other hand, if $ \epsilon_S - \epsilon_D > U$, then the energy can be lowered by 
doubly occupying one of the least energetic singly occupied states
with an electron from the most energetic singly occupied state, as the energy
penalty $U$ for this process is smaller than the energy gain $\epsilon_S - \epsilon_D$ 
in kinetic energy. Again, the original state therefore cannot be the ground state of the system.

From the condition $\epsilon_S = \epsilon_D + U$ it is easy to see that, for given density
$\rho$, the phase HSD can only exist if $U$ is smaller than a certain critical interaction $U_c ( \rho )$: Suppose we increase $U$ starting from $U < U_c ( \rho )$. In order to maintain
$\epsilon_S = \epsilon_D + U$, electrons will leave doubly occupied states  to occupy previously empty states. This can be continued until one of two things happens:
either the interaction has become so strong that there are no doubly occupied states left
(which is equivalent with the statement that $\epsilon_D$  is smaller than the minimal value
of $\epsilon_{\bd{k}}$), or there are no empty state left to be occupied (which is equivalent with the statement that $\epsilon_S$ is larger than the maximal value of $\epsilon_{\bd{k}}$).
In both cases,  there are no longer two Fermi surfaces but one. 
The functional form of $U_c ( \rho )$ depends on the dimensionality and on the
specific form of the energy dispersion $\epsilon_{\bd{k}}$. For nearest neighbor hopping $t > 0$ in one dimension where $\epsilon_{\bd{k}} = - 2 t \cos ( k_x a )$ and $a$ is the lattice spacing the function $U_c ( \rho)$ can be calculated  analytically \cite{Vitoriano00},
 \begin{equation}
  U_c ( \rho ) = W \sin^2 \left( \frac{\pi}{2} \rho \right),
  \label{eq:Uc1d}
  \end{equation}  
where $W = 4 t$ is the bandwidth and the density $\rho$ is normalized such that $ 0 < \rho < 2$ with $\rho =1$ corresponding to a half filled lattice.

Finally, we show that in phase HSD the system is metallic using the following  gap criterion~\cite{Hatsugai92,Gebhard97}:  Let $E_0 ( N )$ be the ground state energy of the system with $N$ particles and define
the chemical potentials 
 \begin{subequations}
 \begin{align}
 \mu_{+}  & = E_0 ( N+1) - E_0 ( N ),
 \\
 \mu_{-} & = E_0 ( N ) - E_0 ( N-1).
 \end{align}
 \end{subequations}
The excitation gap to charge carrying states is then given by \cite{Hatsugai92,Gebhard97}
 \begin{equation}
 \Delta = \mu_+ - \mu_- = E_0 ( N+1) - 2 E_0 ( N ) + E_0 ( N-1).
 \end{equation} 
The system is insulating if $\Delta > 0$. For the phase HSD the chemical potentials are amibguous because this phase has two Fermi surfaces $\bd{k}_D$ and $\bd{k}_S$, defined by $\epsilon_{\bd{k}_D} = \epsilon_D$ and
$\epsilon_{\bd{k}_S} = \epsilon_S$. The chemical potential for adding an extra electron to the $N$ particle system in the phase HSD is therefore
 \begin{equation}
 \mu_+ = \left\{ \begin{array}{ll} \epsilon_{\bd{k}_S + \delta \bd{k} } &
  \mbox{if the electron occupies a hole} \\
   & \mbox{with momentum $\bd{k}_S + \delta \bd{k} $,} \\
  \epsilon_{\bd{k}_D + \delta \bd{k} } + U & \mbox{if the electron doubly occupies  a}
  \\
  & \mbox{state with momentum
   $\bd{k}_D + \delta \bd{k} $.}  
  \end{array} \right.
  \end{equation}
Here $\bd{k}_D + \delta \bd{k}$ and 
$\bd{k}_S + \delta \bd{k}$ is the momemtum of the extra electron which is added close to the
Fermi surface $\bd{k}_D$ or $\bd{k}_S$.
On the other hand, the chemical potential for adding an extra electron to the
$N-1$ particle system is 
\begin{equation}
 \mu_- = \left\{ \begin{array}{ll} \epsilon_{\bd{k}_S  } &
  \mbox{if the electron occupies a hole} \\
   & \mbox{with momentum $\bd{k}_S  $,} \\
  \epsilon_{\bd{k}_D  } + U & \mbox{if the electron doubly occupies a}
  \\
  & \mbox{state with momentum
   $\bd{k}_D  $.}  
  \end{array} \right.
  \end{equation}
Using the continuity of the dispersion and the fact that in this phase $ \epsilon_S - 
\epsilon_D = U$ we conclude that $\Delta =0$, i.e. the excitations have vanishing energy and hence the system is metallic.

\subsubsection{Phase HS: holes and singly occupied states}

In the phase HS the ground state is composed of holes and singly occupied momentum states,
as illustrated in Fig.~\ref{fig:phaseHSD} (b). This phase can also be characterized by the absence of doubly occupied states.
In momentum space the singly occupied and empty states are separated by
a single Fermi surface defined by $\epsilon_{\bd{k}_S }=  \epsilon_S$. 
The ground state in this phase is then explicitly given by
 \begin{equation}
 | E_0^{\rm{HS}}, \{ \sigma_{\bd{k}} \} ) = \prod_{ \epsilon_{\bd{k}} < \epsilon_S } 
 c^{\dagger}_{\bd{k} \sigma_{\bd{k}} } | 0 \rangle. 
 \end{equation}
Note that phase exists only for $\rho < 1$  because for higher densities  double occupancies are needed to accommodate all electrons. 
In order to avoid double occupancies in this phase, the interaction must be larger than the
critical interaction, $U > U_c ( \rho )$.  
When a new particle is added to the system in this phase, it will occupy the least energetic
empty state. The chemical potentials are therefore
 \begin{subequations}
 \begin{align}
 \mu_+ & = \epsilon_{\bd{k}_S + \delta \bd{k}} ,
 \\
 \mu_- & = \epsilon_{\bd{k}_S }.
 \end{align}
 \end{subequations}
Continuity of the dispersion implies
that in the thermodynamic limit $\Delta = \mu_+ - \mu_{-} =0$ so that the system is metallic in this phase and the chemical potential is $\mu = \epsilon_S$.

\subsubsection{Phase SD:  singly and doubly occupied states}

In the phase SD the ground state is an antisymmetrized superposition of singly and doubly occupied momentum states, i.e. there are no holes. Explicitly, the ground state is 
\begin{equation}
 | E^{\rm{SD}}_0 , \{ \sigma_{\bd{k}} \} \rangle =
  \left(  \prod_{ \epsilon_D < \epsilon_{\bd{k}} }c^{\dagger}_{\bd{k}
    \sigma_{\bd{k}} } \right) 
    \left(  \prod_{  \epsilon_{\bd{k}} < \epsilon_D } c^{\dagger}_{\bd{k}
    \uparrow } c^{\dagger}_{\bd{k} \downarrow } \right)
   | 0 \rangle ,
  \end{equation}
where $\epsilon_D$ is the energy of the least energetic singly occupied states. The system then has a single Fermi surface $\bd{k}_D$ defined by $\epsilon_{\bd{k}_D} = \epsilon_D$ which separates singly occupied from doubly occupied states.
A schematic representation of the ground state configuration and the associated momentum distribution in one dimension is shown in Fig.~\ref{fig:phaseHSD}~(c).
Because all states are at least singly occupied,  this phase can only be realized for $\rho > 1$.
Moreover, the interaction must be larger than $ U_c ( \rho )$ to avoid holes. 
Adding a new particle in this phase will double occupy the least energetic singly occupied state. The chemical potentials of the system with $N$ and $N-1$ particles are therefore
  \begin{subequations}
 \begin{align}
 \mu_+ & = \epsilon_{\bd{k}_D + \delta \bd{k}} + U ,
 \\
 \mu_- & = \epsilon_{\bd{k}_D } + U.
 \end{align}
 \end{subequations}
Again, the excitation gap $\Delta = \mu_+ - \mu_-$ vanishes in the thermodynamic limit so that the system is metallic with chemical potential
$\mu = \epsilon_D + U$.

\subsubsection{Phase S: all states singly occupied (Mott insulator)}
In this phase all sites of the reciprocal lattice are singly occupied so that the density is $\rho =1$ and the ground state is
\begin{equation}
 | E^{\rm{S}}_0 , \{ \sigma_{\bd{k}} \} \rangle =
    \prod_{ \bd{k} } c^{\dagger}_{\bd{k}
    \sigma_{\bd{k}} }
   | 0 \rangle ,
  \end{equation}
where the product is over all momenta in the first Brillouin zone. This phase is only stable if
double occupancies are punished by a sufficiently strong interaction
$U > U_c ( 1 )$. Note that $U_c ( 1 ) = W$ is given by the bandwidth, which is the difference in energy between the most and least energetic particles.   
A schematic representation of the ground state configuration and the associated momentum distribution in this phase in one dimension is shown in Fig.~\ref{fig:phaseHSD}~(d).
To see that in this phase the system is a Mott insulator  we consider again the chemical potentials $\mu_{+}$ and $\mu_{-}$. When we add an electron to the system, the least energetic state, which has energy $- W/2$, becomes  doubly occupied so that
 \begin{equation}
  \mu_+ = - \frac{W}{2} + U.
  \label{eq:muplusS}
 \end{equation}
On the other hand, the ground state of the system with $N-1$ particles has a  hole in the most energetic momentum state which, when singly occupied, has energy $W/2$ and hence
\begin{equation}
  \mu_- =  \frac{W}{2}.
  \label{eq:muminusS} 
 \end{equation}
The excitation gap  in the  phase $S$ is therefore
 \begin{equation}
  \Delta = \mu_+ - \mu_- = U - W = U - U_c ( 1) > 0,
  \end{equation}
so that the system is insulating.

\subsection{Phase diagram as a function of density and interaction}

To derive the phase diagram of the model as a function of density $\rho$ and interaction $U$ we need to calculate  the critical interaction $U_c ( \rho )$ separting the phases discussed above. 
Consider first the case $\rho < 1$ where the ground state is either in phase HSD or in HS. The critical interaction $U_c ( \rho )$ is the smallest interaction for which there are no doubly occupied states. As shown in the Appendix, at zero temperature the average occupation of a state with momentum $\bd{k}$ is
 \begin{equation}
 \bar{n}_{\bd{k}} = 
 \Theta ( \mu - \epsilon_{\bd{k}} ) + \Theta ( \mu - \epsilon_{\bd{k}} - U ).
 \label{eq:barn}
 \end{equation}
Here the  chemical potential $\mu = \mu ( \rho , U )$  as a function of density $\rho$ and
interaction $U$ is determined by the implicit equation
 \begin{equation}
 \rho = \frac{1}{ \cal{N}} \sum_{\bd{k}} \bar{n}_{\bd{k}},
 \label{eq:rhomu}
 \end{equation}
where  ${\cal{N}}$ is the number of lattice sites.
To exclude double occupancies the argument of the second $\Theta$-function in
Eq.~(\ref{eq:barn}) must be negative for all momenta $\bd{k}$, i.e.,
 \begin{equation}
 \mu - \epsilon_{\bd{k}} < U, \; \; \; \mbox{for all $\bd{k}$}.
 \label{eq:cond1}
 \end{equation}
The left-hand side of this expression  is maximal if we replace $\epsilon_{\bd{k}}$ by its lowest possibly values $- W/2$, where $W$ is the total bandwidth. 
The smallest interaction where  the condition (\ref{eq:cond1}) is satisfied is therefore
given by the solution of the implicit equation
 \begin{equation}
 U_c ( \rho ) = \mu ( \rho, U_c ( \rho )) + W/2.
 \label{eq:Uc1}
 \end{equation}
Note that for $\rho \rightarrow 1$ where $\mu = W/2$, this implies $U_c (1) = W$ independent of the precise form of the energy dispersion. 

Next, consider the case $\rho > 1$  where $U_c ( \rho )$ is defined by the minimal interaction for which there are no holes, which is the case if
 \begin{equation}
 0 \leq \mu - \epsilon_{\bd{k}}, \; \; \; \mbox{for all $\bd{k}$}.
 \label{eq:cond2}
 \end{equation}
The right-hand side of this inequality is minimal if $\epsilon_{\bd{k}}$ assumes its maximum
$W/2$, so that $U_c ( \rho )$  is for $\rho > 1$ determined by
 \begin{equation}
 \mu ( \rho, U_c ( \rho)) = W/2.
 \label{eq:Uc2}
 \end{equation}

The disadvantage of Eqs.~(\ref{eq:Uc1}) and (\ref{eq:Uc2}) is that these implicit equations require the chemical  potential  $\mu = \mu ( \rho , U )$ as function of density, which can be obtained by 
inverting Eq.~(\ref{eq:rhomu}). In practice, it is more convenient to derive an implicit relation between $\rho$ and $U_c$ directly from Eq.~(\ref{eq:rhomu}). 
Consider first the phase HS which is realized for $U > U_c (\rho)$ and $\rho < 1$. 
The particle density can then be written as
 \begin{equation}
 \rho = \int_{ - W/2}^{ \epsilon_S} d \epsilon \nu ( \epsilon ),
 \end{equation}
where $\epsilon_S$ is the energy of the highest singly occupied state and we have introduced 
the density of states
 \begin{equation}
  \nu ( \epsilon ) = \frac{1}{\cal{N}} \sum_{\bd{k}} \delta ( \epsilon - \epsilon_{\bd{k}} ).
  \end{equation}
As discussed in Sec.~\ref{subsubsec:ESD}, if we lower the value of $U$, then
double occupancies will emerge at the interaction $U_c$ where the energetic difference between the most and least energetic singly occupied state is given by  $U_c$, i.e.,
 \begin{equation}
 \epsilon_S - ( - W/2) = U_c.
 \end{equation}
Hence, for $\rho < 1$ the critical interaction $U_c ( \rho )$ is given by the implicit equation 
 \begin{equation}
 \rho = \int_{ - W/2}^{ U_c - W/2} d \epsilon \nu ( \epsilon ).
 \label{eq:rhoUc}
 \end{equation}
For nearest-neighbor hopping in one dimension the integration  
in Eq.~(\ref{eq:rhoUc}) can be carried out exactly and the resulting equation can be explicitly solved for $U_c = U_c ( \rho )$; see Eq.~(\ref{eq:Uc1d}).

Next, consider the regime $U >U_c ( \rho)$ for $\rho > 1$ where the ground state  SD has no holes.
The particle density is then given by
 \begin{equation}
 \rho = 2 \int_{ - W/2}^{\epsilon_D} d \epsilon \nu ( \epsilon ) + 
\int_{ \epsilon_D}^{W/2} d \epsilon \nu ( \epsilon ),
 \end{equation}
where $\epsilon_D$ is the energy of lowest singly occupied state. 
At the critical interacton $U_c ( \rho )$
the difference in the kinetic energies of the most and least energetic singly occupied states is exactly $U_c$, i.e.,
 \begin{equation}
  W/2 - \epsilon_D = U_c.
  \end{equation}
Our implicit relation between $\rho$ and $U_c $ is therefore
\begin{equation}
 \rho = 2 \int_{ - W/2}^{W/2 - U_c} d \epsilon \nu ( \epsilon ) + 
\int_{ W/2 - U_c}^{W/2} d \epsilon \nu ( \epsilon ).
 \label{eq:rhoUc2}
 \end{equation} 
For nearest-neighbor hopping in one dimension this gives exactly the same relation
(\ref{eq:Uc1d}) as for $\rho < 1$. In fact, if we assume $\nu ( \epsilon ) = \nu ( - \epsilon )$ and use $\int_{ - W/2}^{W/2} d \epsilon \nu ( \epsilon ) =1$, we can rewrite 
Eq.~(\ref{eq:rhoUc2}) in the form
  \begin{equation}
 2- \rho = \int_{ - W/2}^{ U_c - W/2} d \epsilon \nu ( \epsilon ).
 \label{eq:rhoUc22}
 \end{equation}
Comparing this with Eq.~(\ref{eq:rhoUc}) we see that 
for $\rho > 1$ the function $U_c ( \rho )$ can be obtained from
the corresponding function $U_c^{ <} ( \rho ) = \Theta ( 1 - \rho ) U_c ( \rho)$ for $\rho < 1$ using
$U_c ( \rho ) = U^<_c ( 2 - \rho )$.

\begin{figure}[tb]
 \begin{center}
  \centering
 \includegraphics[width=0.5\textwidth]{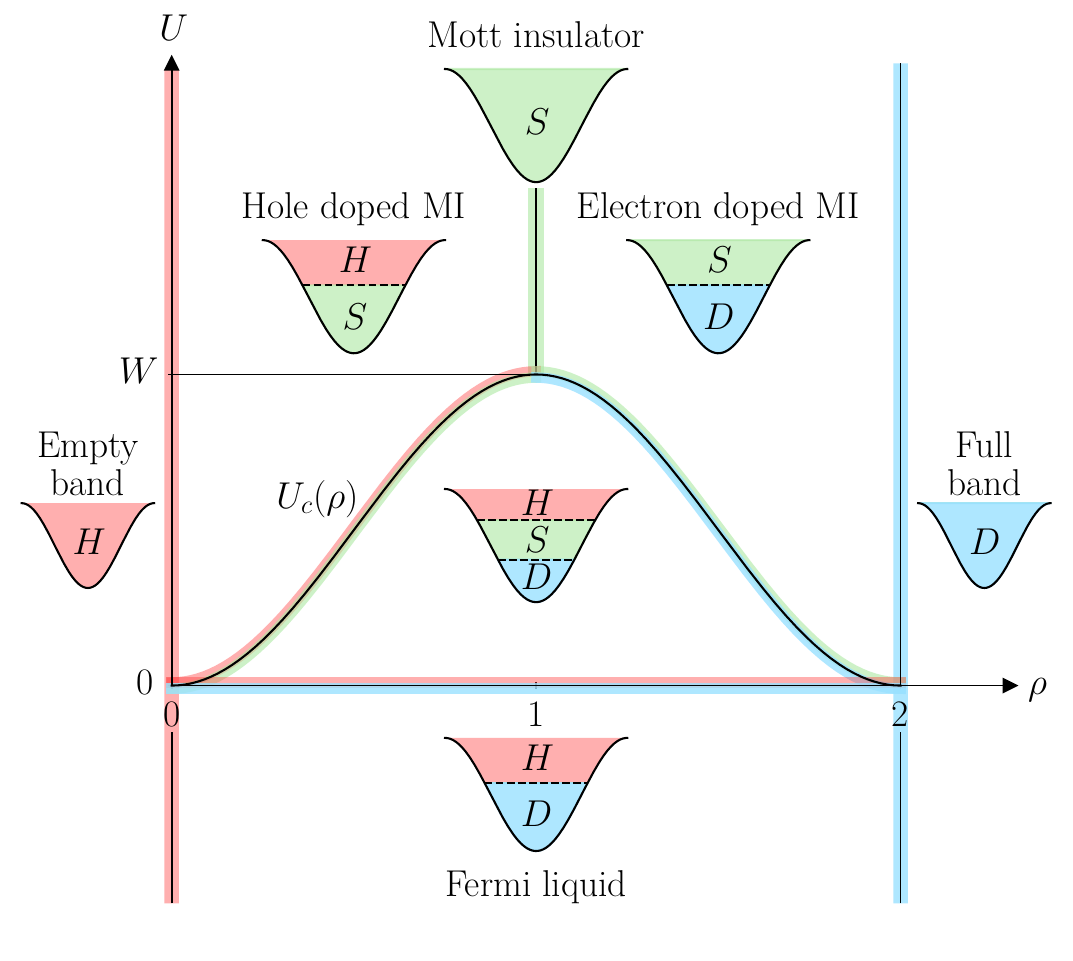}
   \end{center}
  \vspace{-4mm}
  \caption{%
Phase diagram of the HK model as a function of the density $\rho$ and the interaction $U$. The color coding indicates the occupancies of momentum states:  holes (H, red), 
singly occupied states (S, green), and doubly occupied states (D, blue).
The insets represent the corresponding band fillings. 
}
\label{fig:7phases}
\end{figure}

\begin{figure}[htb]
 \begin{center}
  \centering
 \includegraphics[width=0.5\textwidth]{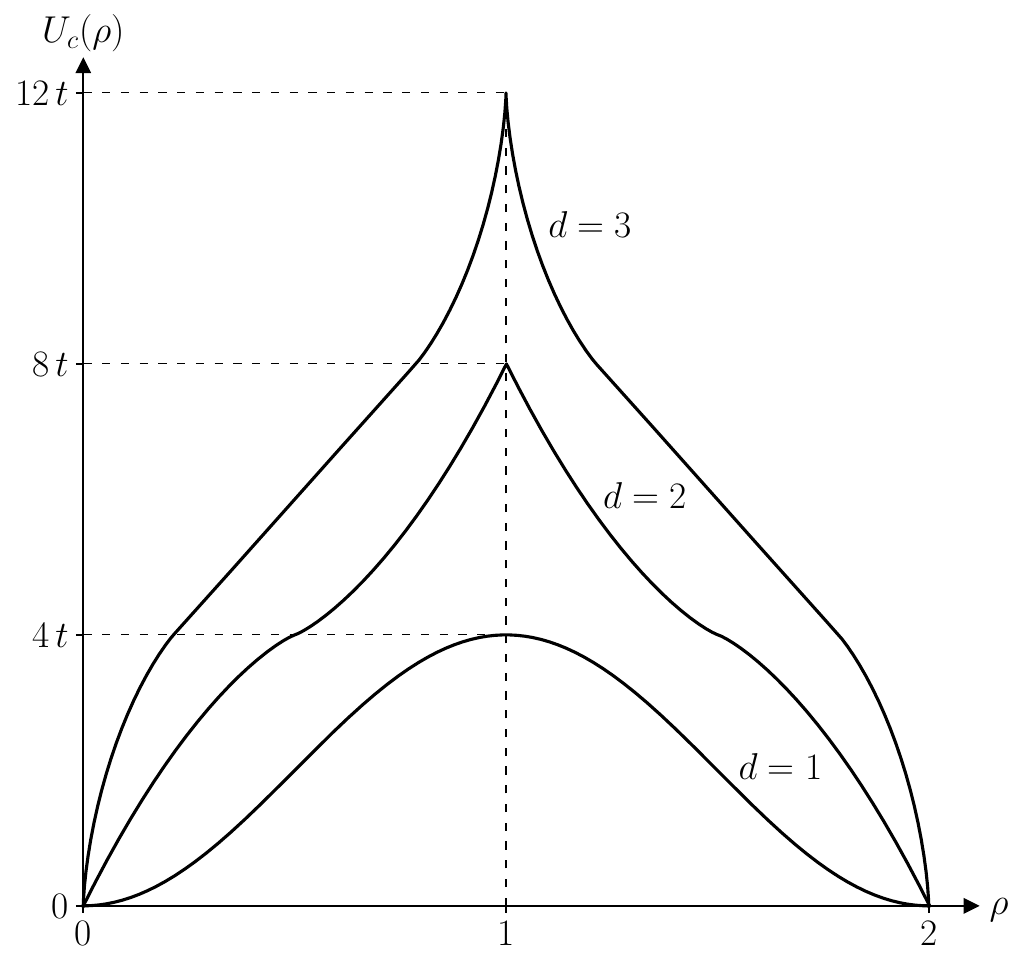}
   \end{center}
  \vspace{-4mm}
  \caption{%
Critical interaction $U_c ( \rho )$ separating the HSD phase from the HS phase and the SD phase 
for a $d$-dimensional tight-binding dispersion of the type (\ref{eq:epsdef})
in $d=1,2,3$.
}
\label{fig:Uc123}
\end{figure}

To conclude this subsection, let us explicitly draw the ground state phase diagram  of the HK model
as a function of density and interaction
assuming nearest neighbor hopping $t > 0$ on a $d$-dimensional hypercubic lattice with lattice spacing $a$. The corresponding energy dispersion  is
 \begin{equation}
  \epsilon_{\bd{k}} = - 2 t \sum_{ \mu =1}^d \cos ( k_\mu a ),
  \label{eq:epsdef}
  \end{equation}
implying $W = 4 d t$ for the bandwidth.
In one dimension the critical interaction $U_c ( \rho )$ can be calculated analytically and is explicitly given in Eq.~(\ref{eq:Uc1d}). The resulting phase diagram is shown in 
Fig.~\ref{fig:7phases}. 
In  $d > 1$ dimensions the relevant equations for $U_c ( \rho)$ cannot be solved analytically. However, the phase boundaries  for $d > 1$
can easily be obtained numerically. 
In Fig.~\ref{fig:Uc123}  we show the function $U_c ( \rho )$ 
for the tight-binding dispersion (\ref{eq:epsdef})
in $d=1,2,3$.

Note that the ground state in all phases containing singly occupied states is degenerate because the spin-projections of the electrons residing on the singly occupied reciprocal lattice sites can assume two possible values $\sigma_{\bd{k}} = \uparrow, \downarrow$.
In Sec.~\ref{sec:pseudospin} we shall write down a minimally modified version of the
HK model which has a unique ground state for all densities.

\subsection{Phase diagram as a function of bandwidth and chemical potential in units of $U$}

\begin{figure}[tb]
 \begin{center}
  \centering
 \includegraphics[width=0.5\textwidth]{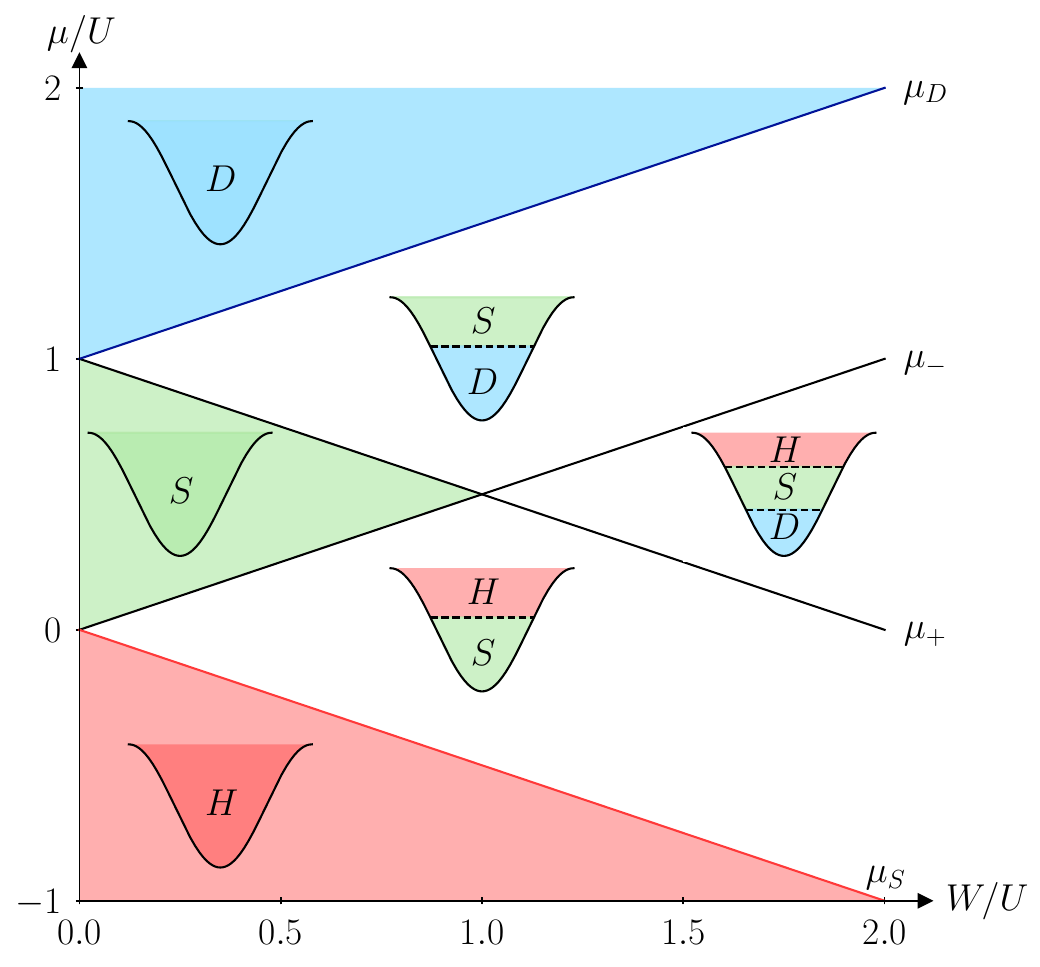}
   \end{center}
  \vspace{-4mm}
  \caption{%
Phase diagram of the HK model as a function of the chemical potential $\mu/U$ and
band width $W/U$ in units of the interaction $U$.
The thick black lines represent the 
chemical potentials
$\mu_+$, $\mu_-$ defined in Eq.~(\ref{eq:mupmdef}). The thick red lines 
represents $\mu_S = - W/2$, while the thick blue line represents $\mu_D = U + W/2$.
}
\label{fig:phasesmu}
\end{figure}

For completeness, we also derive the phase diagram of the HK model as a function of 
chemical potential 
$\mu /U$ and bandwidth $W/U$ in units of the interaction $U$, which turns out to be independent of the dimensionality \cite{Vitoriano00}.
To this end we  note that the chemical potentials $\mu_{+}$ and $\mu_-$
in Eqs.~(\ref{eq:muplusS}) and (\ref{eq:muminusS}) represent the boundaries separating the insulating Mott phase from the metallic phases. Dividing all energies by $U$, the boundaries are 
 \begin{subequations}
 \label{eq:mupmdef}
 \begin{align}
 \frac{\mu_+}{U} & = - \frac{W}{2U} + 1,
 \\
 \frac{\mu_-}{U} & = \frac{W}{2U},
 \end{align}
 \end{subequations}
which are represented by  thick black lines in Fig.~\ref{fig:phasesmu}. 
If $\mu_- < \mu < \mu_+$ there is a finite gap $\Delta = \mu_+ - \mu_{-}$ between  the chemical potentials of the system with $N$ and $N-1$ particles, so that the  system is a Mott insulator. This Mott regime is represented by the green shaded triangle S 
(singly occupied states) 
in  Fig.~\ref{fig:phasesmu}, while the red line represents the chemical potential
$\mu_S = - W/2$ corresponding to the least energetic occupied state;
for chemical potentials below this line the system is completely empty.
Finally, the blue  line represents $\mu_D = U + W/2$  corresponding to the largest possible energy of  an electron; for $ \mu > \mu_D$ the system is completely filled, i.e., all states are doubly occupied.
To identify the phases HSD, HS, and SD in the phase diagram, we note that doubly occupied states exist only if $\epsilon_D > - W/2$, while holes exist only if
$\epsilon_S < W/2$. But in phase HSD we have $\mu = \epsilon_S = \epsilon_D + U$.
We conclude that for
 \begin{align}
 \mu > U -W/2 & = \mu_+ \; \; \; \begin{array}{l} \mbox{the system contains}  \\
  \mbox{doubly occupied states,} \end{array}
 \end{align}
while for
\begin{align}
 \mu < W/2 & = \mu_- \; \; \; \begin{array}{l} \mbox{the system contains holes.}  \\
 \end{array}
 \end{align}
In phase HSD both conditions must be fulfilled, corresponding to $\mu_+ < \mu < \mu_-$. On the other hand, the phase HS is characterized  by the absence of doubly occupied states and by the presence of holes, corresponding to $\mu < \mu_+$ and  $ \mu < \mu_-$. Finally, in the phase SD the system has no holes  but doubly occupied states, corresponding to $\mu > \mu_+$ and $\mu > \mu_-$. Taking into account all inequalities discussed above, we arrive
at the phase diagram shown in Fig.~\ref{fig:phasesmu}, which agrees with the phase diagram given 
in Ref.~[\onlinecite{Vitoriano00}].

\section{Pseudospin picture and non-degenerate HK model}
\label{sec:pseudospin}

In the spirit of 
Andersons pseudospin formulation of BCS theory \cite{Anderson58}, in this section we rewrite the HK model in terms of suitably defined  pseudospin operators, which allows us to interpret the phases of the model in terms of the magnetic phases  
of pseudospin model in reciprocal space, where each site of the reciprocal lattice hosts two Ising spins. We also show that within the pseudospin formalism it is straightforward to write down
an exactly solvable modification of  HK model which has a unique ground state for all densities.

\begin{figure}[tb]
 \begin{center}
  \centering
 \includegraphics[width=0.5\textwidth]{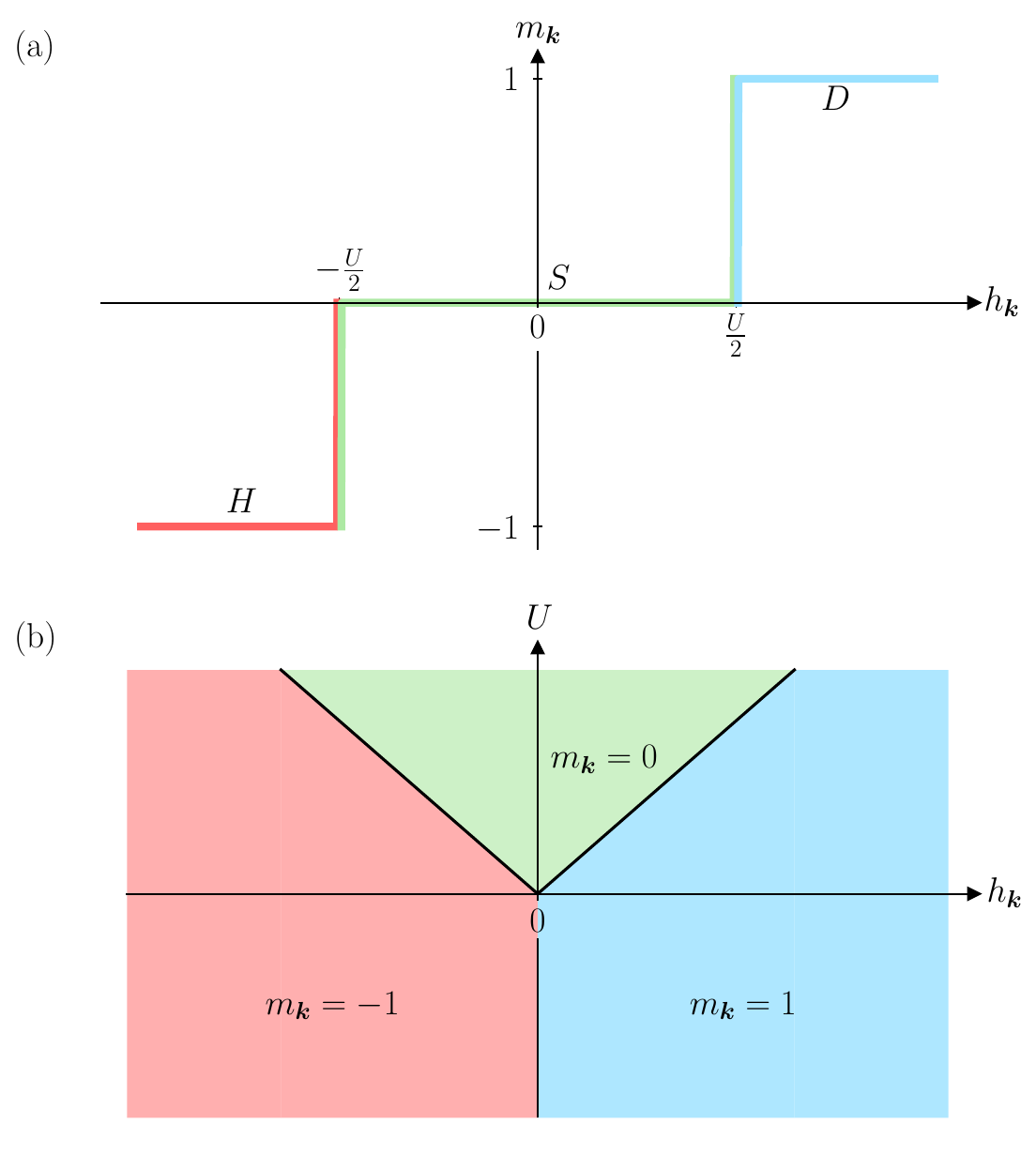}
   \end{center}
  \vspace{-4mm}
  \caption{%
(a) Local magnetic moment $m_{\bd{k}} = \langle S^z_{\bd{k}} \rangle$
of the pseudospin HK model (\ref{eq:HKpseudo}) as function of the
pseudomagnetic field $h_{\bd{k}}$. (b) 
Phase diagram of the HK model in the plane spanned by $h_{\bd{k}}$ and $U$.}
\label{fig:mag}
\end{figure}

Because for each momentum $\bd{k}$ and spin projection $\sigma$ the
occupation  number operator $n_{\bd{k} \sigma}$ has eigenvalues $0$ and $1$, 
the operator
\begin{equation}
S^z_{\bd{k} \sigma} = n_{\bd{k} \sigma} - 1/2
 \label{eq:Szdef}
 \end{equation}
has eigenvalues $\pm 1/2$ and can be identified with the $z$-component of a
pseudospin-$1/2$ operator $\bd{S}_{\bd{k} \sigma}$.  
To emphasize the relation between  the HK model and an effective spin model,  
it is convenient to replace the original HK Hamiltonian (\ref{eq:HK}) by
 \begin{align}
 {\cal{H}}_{\rm HK}^\prime &  = 
 \sum_{\bd{k}}  \biggl[ \epsilon_{\bd{k}}  (
 n_{ \bd{k} \uparrow} + n_{\bd{k} \downarrow} -1 )
 \nonumber
 \\ 
 &  \hspace{7mm} + U  \left( n_{\bd{k} \uparrow} - \frac{1}{2} \right) \left( n_{\bd{k} \downarrow} 
 - \frac{1}{2} \right) - \frac{U}{4} \biggr]
 \nonumber
 \\
 & =  \sum_{\bd{k}}  \biggl[ \epsilon_{\bd{k}}  (
 S^z_{ \bd{k} \uparrow} + S^z_{\bd{k} \downarrow} )
 + U  \left( S^z_{\bd{k} \uparrow}  S^z_{\bd{k} \downarrow} - \frac{1}{4} \right)  \biggr]
 ,
 \label{eq:HKph}
 \end{align}
which can be obtained from the original HK Hamiltonian (\ref{eq:HK})  by shifting the 
dispersion $\epsilon_{\bd{k}}$ by a constant (which can be absorbed  by re-defining 
the chemical potential) and adding an overall constant.
The subtraction $- U /4$ in Eq.~(\ref{eq:HKph}) 
is necessary to assign vanishing interaction energy to empty states, which is  convenient to emphasize the symmetry between empty and doubly occupied states.
In the occupation number basis defined in Eq.~(\ref{eq:basisoccupation})
the Hamiltonian (\ref{eq:HKph}) is then a direct sum of the following $4 \times 4$ blocks labelled by
the momentum $\bd{k}$, 
\begin{equation}
 {\mathbf{H}}^{\prime}_{\bd{k}} =
  \left( \begin{array}{cccc} - \epsilon_{\bd{k}}  & 0 & 0 & 0 \\
  0& - \frac{U}{2} & 0 &  0\\
  0 & 0 & - \frac{U}{2}  & 0 \\
  0 &  0 & 0 &  \epsilon_{\bd{k}}  \end{array}
  \right).
  \label{eq:HKmatrix2}
  \end{equation}  
Note that this matrix can be obtained from the matrix
${\mathbf{H}}_{\bd{k}}$ defined in Eq.~(\ref{eq:HKmatrix1}) by shifting
$ \epsilon_{\bd{k}} \rightarrow \epsilon_{\bd{k}} -   U/2$ and then subtracting $\epsilon_{\bd{k}}$ from all diagonal elements.
It is convenient to define the corresponding grand canonical
Hamiltonian  as follows,
\begin{align}
 & \tilde{\cal{H}}_{\rm HK}  = {\cal{H}}_{\rm HK}^\prime - \mu \sum_{\bd{k} \sigma} 
 \left(  n_{\bd{k} \sigma}  -\frac{1}{2} \right)
 \nonumber
 \\
 & =  \sum_{\bd{k}}  \biggl[ - h_{\bd{k}}  (
 S^z_{ \bd{k} \uparrow} + S^z_{\bd{k} \downarrow} ) 
 + U  \left( S^z_{\bd{k} \uparrow}  S^z_{\bd{k} \downarrow} - \frac{1}{4} \right)  \biggr],
  \label{eq:HK3}
 \end{align}
where 
 \begin{equation}
 h_{\bd{k}} = \mu -  \epsilon_{\bd{k}}
 \end{equation}
plays the role of an inhomogeneous pseudomagnetic field in momentum space.
Note that we have replaced the usual chemical potential term by
$\mu \sum_{\bd{k} \sigma}  ( n_{\bd{k} \sigma} - 1/2)$ so that we can express our  model in terms of two flavors of pseudospin operators $S^{z}_{\bd{k} \uparrow}$
and $S^{z}_{\bd{k} \downarrow}$ labelled by the momentum and the physical spin $\sigma$.
Finally, introducing the total $z$-component of the pseudospin,
 \begin{equation}
  S^z_{\bd{k}} = S^z_{\bd{k} \uparrow} + S^z_{\bd{k} \downarrow},
  \label{eq:Stotdef}
  \end{equation}
and  using the identity
 \begin{align}
  S^z_{\bd{k} \uparrow}  S^z_{\bd{k} \downarrow} & = 
  \frac{1}{2} 
  \left[ (S^z_{\bd{k}} )^2 -  (S^z_{\bd{k} \uparrow} )^2 -
  (S^z_{\bd{k} \downarrow} )^2 \right]
  \nonumber
  \\
  & = \frac{1}{2} 
  \left[ (S^z_{\bd{k}} )^2 -  \frac{1}{2} \right],
  \end{align}
our pseudospin HK Hamiltonian  can be written as
\begin{align}
 \tilde{\cal{H}}_{\rm HK} & = \sum_{\bd{k}} \tilde{\cal{H}}_{\bd{k}}  
 = \sum_{\bd{k}}  \left\{ - h_{\bd{k}}  
 S^z_{ \bd{k}} 
 + \frac{U}{2}  \left[  ( S^z_{\bd{k}} )^2  - 1 \right] \right\}.
 \label{eq:HKpseudo}
 \end{align}
Note that holes (i.e., empty states) in the original particle picture correspond to
states with $S^z_{\bd{k}} = -1$ in the pseudospin picture, while doubly occupied states
in the particle picture correspond to states with maximal spin $S^z_{\bd{k}} =+1$.
In the occupation number basis~(\ref{eq:basisoccupation}) the 
blocks $\tilde{\cal{H}}_{\bd{k}}  $ are represented by
\begin{equation}
 \tilde{\mathbf{H}}_{\bd{k}} =
  \left( \begin{array}{cccc} h_{\bd{k}}  & 0 & 0 & 0 \\
  0& - \frac{U}{2} & 0 &  0\\
  0 & 0 & - \frac{U}{2}  & 0 \\
  0 &  0 & 0 &  - h_{\bd{k}}  \end{array}
  \right).
  \label{eq:HKmatrix3}
  \end{equation}  
The pseudospin picture makes the competition between kinetic and potential energy responsible for the Mott transition evident: the Zeeman energy $- h_{\bd{k}} S^z_{\bd{k}}$ associated with the pseudomagnetic field tends to align the spins in the field direction 
and hence favors states with finite magnetization, while the
interaction energy $\frac{U}{2} ( ( S_{\bd{k}}^z )^2 -1 )$ punishes spin configurations with finite local magnetic moments. 
The phases HSD, HS, SD, and S discussed in
Sec.~\ref{sec:phasesHK} can be interpreted as phases involving different types of magnetic domain walls in momentum space. 
As shown in the Appendix, for vanishing temperature the
magnetization $m_{\bd{k}} = \langle S^z_{\bd{k}} \rangle$ of the
pseudospin HK model (\ref{eq:HKpseudo}) as function of the pseudomagnetic field
$h_{\bd{k}}$ is
given by
\begin{align}
  m_{\bd{k}}  & = 
   \Theta ( | h_{\bd{k}} | - U/2 ) {\rm sign } h_{\bd{k}}
  \nonumber
  \\
  &  = \left\{ 
 \begin{array}{cc}  
  0  & \mbox{if $| h_{\bd{k}} | < U/2$},\\
 {\rm sign } h_{\bd{k}} & \mbox{if $ | h_{\bd{k}} | > U/2$},
\end{array} \right. 
 \label{eq:mkres}
  \end{align}
which is illustrated  in Fig.~\ref{fig:mag}.
The relevant range of the pseudomagnetic field $h_{\bd{k}} = \mu - \epsilon_{\bd{k}}$
depends in the bandwidth and the chemical potential (or the density). The various phases
discussed in Sec.~\ref{sec:phasesHK} can be realized by selecting different intervals of the
magnetization curve shown in Fig.~\ref{fig:mag}. In particular, in the phase HSD
the relevant interval of $h_{\bd{k}}$ covers both domain walls where the magnetization exhibits discontinuities, while in the Mott insulating phase $S$ only the central piece with $m_{\bd{k}} =0$ is relevant. Note that in the pseudospin picture the
Mott insulator corresponds to a non-magnetic state where $\langle S^z_{\bd{k}} \rangle =0$ for all $\bd{k}$ in spite
of the presence of a pseudomagnetic field $h_{\bd{k}}$. This corresponds to a magnetization plateau at zero magnetization, which can exist due to the gap in the energy spectrum.

The ground states in all phases of the HK model involving reciprocal lattice sites
with $m_{\bd{k}} =0$ are degenerate
because a state with $S^z_{\bd{k}} = S^z_{\bd{k} \uparrow} + S^z_{\bd{k} \downarrow} =0$ can be realized either by choosing $S^{z}_{\bd{k} \uparrow} = 1/2 $ and 
$S^{z}_{\bd{k} \downarrow} = - 1/2 $, or
  $S^{z}_{\bd{k} \uparrow} = -1/2 $ and 
$S^{z}_{\bd{k} \downarrow} =  1/2 $. 
We can now write down a simple generalization of the
HK model which does not exhibit any ground state degeneracies.
In the pseudospin picture it is obvious how to lift this degeneracy without destroying the
symmetry  or the exact solubility of the Hamiltonian:  
we simply add a term involving the square of the total pseudospin 
$\bd{S}_{\bd{k}} = \bd{S}_{\bd{k} \uparrow} + \bd{S}_{\bd{k} \downarrow}$, 
so that our modified HK model in pseudospin notation is
 \begin{equation}
 \tilde{\cal{H}}^V_{\rm HK}  =  \sum_{\bd{k}} \tilde{\cal{H}}^V_{\bd{k}},
 \label{eq:HKS}
 \end{equation}
with
 \begin{equation}
 \tilde{\cal{H}}^V_{\bd{k}} = - h_{\bd{k}} 
 S^z_{ \bd{k}}
 + \frac{U}{2}   \left[  ( S^z_{\bd{k} })^2    - 1 \right] +  \frac{V}{2} (
 \bd{S}_{\bd{k}}^2 -2) .
 \label{tildeHk}
  \end{equation}  
Here $\bd{S}_{\bd{k}}^2 = ( S^z_{\bd{k}} )^2
+ \frac{1}{2} ( S^+_{\bd{k}} S^-_{\bd{k}} + S^-_{\bd{k}}
 S^+_{\bd{k}} )$ is the square of the total pseudospin, and the subtraction in
$ \bd{S}_{\bd{k}}^2 -2$ is introduced for convenience to remove the energy shift of triplet states induced by the interaction $V$.
Because for all $\bd{k}$ the operators $ \bd{S}_{\bd{k}}^2 $ and 
$S^z_{\bd{k}}$ commute, we may diagonalize $\tilde{\cal{H}}^V_{\bd{k}}$
in the simultaneous  eigenbasis of  $ \bd{S}_{\bd{k}}^2 $ and 
$S^z_{\bd{k}}$. Keeping in mind that the total pseudospin
$\bd{S}_{\bd{k}} = \bd{S}_{\bd{k} \uparrow} + \bd{S}_{\bd{k} \downarrow}$ is the sum of two spin-$1/2$ operators,   the eigenstates $| j m )$    of $\tilde{\cal{H}}^V_{\bd{k}}$
can be labelled  by the
total spin quantum numbers $j = 0,1$ of $ \bd{S}_{\bd{k}}^2 $ and by the eigenvalues $m$ of
$S^z_{\bd{k}}$. Explicitly, the eigenstates are
 \begin{subequations}
\begin{align}
 | 1 , - 1 )_{\bd{k}} & = | 0 \rangle,
 \\
 | 0 , 0 )_{\bd{k}} & = \frac{1}{\sqrt{2}}
  \left( c^{\dagger}_{\bd{k} \uparrow}| 0\rangle -  c^{\dagger}_{\bd{k} \downarrow}| 0 \rangle\right),
 \\
 | 1 , 0 )_{\bd{k}} & = \frac{1}{\sqrt{2}}
  \left( c^{\dagger}_{\bd{k} \uparrow}| 0\rangle +  c^{\dagger}_{\bd{k} \downarrow}| 0 \rangle \right),
 \\
 | 1 , 1 )_{\bd{k}} & =  c^{\dagger}_{\bd{k} \uparrow} c^{\dagger}_{\bd{k} \downarrow} | 0 \rangle.
  \end{align}
 \end{subequations}
In this basis the block $\tilde{\cal{H}}^V_{\bd{k}}$ is represented by the diagonal matrix
\begin{equation}
 \tilde{\mathbf{H}}^V_{\bd{k}} =
  \left( \begin{array}{cccc}  h_{\bd{k}}    & 0 & 0 & 0 \\
  0&  - \frac{U}{2} - V   & 0 &  0\\
  0 & 0 & - \frac{U}{2}   & 0 \\
  0 &  0 & 0 &   - h_{\bd{k}}    \end{array}
  \right).
  \label{eq:HKSmatrix}
  \end{equation}
Note that the  states with energies $-U/2- V $ and $- U / 2 $ are both associated with  $S^z_{\bd{k}} =0$. For large $U$ and vanishing total magetization both states describe a Mott insulator, with the ground state given by the singlet state 
$| 0 , 0 )_{\bd{k}}$
 for $V >0$, and by the triplet state  $| 1 , 0 )_{\bd{k}} $   for $V < 0$.

Finally, let us express the modified HK Hamiltonian (\ref{eq:HKS}) in terms of canonical fermions. While the fermionic representation of
$S^z_{\bd{k} \sigma} $ is given in Eq.~(\ref{eq:Szdef}), the proper construction of the fermionic representation of the transverse components $S^{+ }_{\bd{k} \sigma}$ and
$S^-_{\bd{k} \sigma}$ is more subtle. By definition, 
the transverse components 
should be defined such that the usual spin algebra
 \begin{subequations}
 \begin{align}
   [ S^{z}_{\bd{k} \sigma} , S^{\pm}_{\bd{k} \sigma} ] & = \pm S^{\pm}_{\bd{k} \sigma} ,
   \\
   [ S^{+}_{\bd{k} \sigma} , S^{-}_{\bd{k} \sigma} ] & = 2  S^{z}_{\bd{k} \sigma} ,
   \end{align}
   \end{subequations}
is satisfied for all momenta $\bd{k}$ and physical spin projections $\sigma$.
Moreover, all components of distinct spin operators $\bd{S}_{\bd{k} \sigma}$ and
$\bd{S}_{\bd{k}^\prime \sigma^\prime}$ should commute. To implement this,
recall that the fermionic representation of spin-$1/2$ lattice models 
developed by 
Jordan and Wigner \cite{Jordan28,Tsvelik95}
realizes this commutation condition via non-local phase operators.
Given the fact that for the HK model pseudospins with different momenta are completely decoupled, it is sufficient to require that
for $\sigma^{\prime} \neq \sigma$  all components of $\bd{S}_{\bd{k} \sigma}$ and 
$\bd{S}_{\bd{k} \sigma^{\prime}}$ should commute. Because the generalized HK model
(\ref{eq:HKS})
depends on two types of pseudospins $\bd{S}_{\bd{k} \uparrow}$ and $\bd{S}_{\bd{k} \downarrow}$, we cannot avoid Jordan-Wigner phase operators. A possible choice is
 \begin{subequations}
 \begin{align}
 S^+_{\bd{k} \uparrow} & = c^{\dagger}_{\bd{k} \uparrow}   , 
 \\
 S^-_{\bd{k} \uparrow} & = c_{\bd{k} \uparrow},
 \\ 
 S^+_{\bd{k} \downarrow} & = c^{\dagger}_{\bd{k} \downarrow} 
 e^{ i \pi n_{\bd{k} \uparrow}} =   c^{\dagger}_{\bd{k} \downarrow} 
 ( 1 - 2  n_{\bd{k} \uparrow} ) ,
 \\
 S^-_{\bd{k} \downarrow} & = e^{ - i \pi n_{\bd{k} \uparrow} } c_{\bd{k} \downarrow}
=   ( 1 - 2  n_{\bd{k} \uparrow} ) c_{\bd{k} \downarrow}.
 \end{align}
 \end{subequations}
Note that in $S^z_{\bd{k} \sigma} = S^+_{\bd{k} \sigma} S^-_{\bd{k} \sigma} -  1/2$ the phases cancel. Using the operator identities 
$c_{\bd{k} \sigma} ( 1 - 2 n_{\bd{k} \sigma} ) = - c_{\bd{k} \sigma}$, $  (1 - 2 n_{\bd{k} \sigma } ) c^\dagger_{\bd{k} \sigma}  = - c^\dagger_{\bd{k} \sigma}$, and
$n_{\bd{k} \sigma}^2 = n_{\bd{k} \sigma}$,  
the fermionic representation of the
operator $\bd{S}_{\bd{k}}^2 $ in Eq.~(\ref{eq:HKS}) can be written as
  \begin{equation}
  \bd{S}_{\bd{k}}^2 = 2 - ( n_{\bd{k} \uparrow}- n_{\bd{k} \downarrow} )^2 
  + c^{\dagger}_{\bd{k} \uparrow} c_{\bd{k} \downarrow} + 
c^{\dagger}_{\bd{k} \downarrow} c_{\bd{k} \uparrow}.
\end{equation}
Substituting this and $S^z_{\bd{k}} = n_{\bd{k} \uparrow } + n_{\bd{k} \downarrow } -1$
into Eq.~(\ref{eq:HKSmatrix}) we finally obtain
 \begin{align}
 \tilde{\cal{H}}^V_{\bd{k}} & =   h_{\bd{k}}   - \left( h_{\bd{k}} + \frac{ U+V}{2} \right) ( n_{\bd{k} \uparrow} + n_{\bd{k} \downarrow } )  
 \nonumber
 \\
 & +
  ( U+ V ) n_{\bd{k} \uparrow} n_{\bd{k} \downarrow}
  + \frac{V}{2} \left( c^{\dagger}_{\bd{k} \uparrow} c_{\bd{k} \downarrow} + 
c^{\dagger}_{\bd{k} \downarrow} c_{\bd{k} \uparrow} \right).
 \label{eq:HKSfermi}
\end{align}
Without pseudospin notation it would have been rather difficult to guess that
the Hamiltonian (\ref{eq:HKSfermi}) is an exactly solvable deformation of the HK model with a unique ground state.
In the following section we set again $V=0$ and use the pseudospin picture  to investigate the effect of momentum mixing on the ground state of the repulsive HK model.

\section{Fermion condensation due to momentum mixing}
\label{sec:fermicond}

To obtain a simple generalization of the HK model including momentum mixing, 
we add a 
Landau type of  interaction to the HK Hamiltonian. The resulting HKL Hamiltonian
${\cal{H}}_{\rm HKL}$
defined in Eq.~(\ref{eq:HKL}) 
is not exactly  solvable any more. In fact, the complexity of finding the ground state
of the HKL Hamiltonian is comparable to finding the ground state of a generalized Ising model
in an inhomogeneous magnetic field. To see this, 
we use the definition $S^z_{\bd{k}} = S^{z}_{ \bd{k} \uparrow} + S^z_{\bd{k} \downarrow} =
 n_{\bd{k} \uparrow} + n_{\bd{k} \downarrow} -1 $ to write
the Landau interaction in Eq.~(\ref{eq:HKL}) as
 \begin{align}
  & \frac{1}{2 \cal{V}} 
 \sum_{\bd{k} \bd{k}^{\prime} \sigma \sigma^{\prime} } f_{\bd{k}  \bd{k}^\prime }  n_{\bd{k} \sigma} n_{\bd{k}^{\prime} \sigma^{\prime}}
 = \frac{1}{2 \cal{V} } \sum_{\bd{k} \bd{k}^{\prime}}   f_{ \bd{k} \bd{k}^\prime}    S^z_{\bd{k}} S^z_{\bd{k}^\prime} 
 \nonumber
 \\
 &
 + \sum_{\bd{k}} S^z_{\bd{k}} \left(  \frac{1}{\cal{V}}\sum_{\bd{k}^\prime} f_{ \bd{k} \bd{k}^\prime}  \right)
 + \frac{1}{2 \cal{V}} \sum_{\bd{k} \bd{k}^{\prime} } f_{ \bd{k} \bd{k}^\prime}  ,
 \label{eq:landauint}
 \end{align}
Obviously, the quantity $J_{\bd{k} \bd{k}^\prime} = \frac{1}{\cal{V}} f_{\bd{k} \bd{k}^\prime}$
plays the role of an exchange interaction between the composite Ising pseudospins $S^z_{\bd{k}}$ attached to the sites $\bd{k}$ of the reciprocal lattice. 
Dropping the last two terms in Eq.~(\ref{eq:landauint}) (which can be absorbed into a re-definition of the
magnetic field and the total energy), we arrive at the following pseudospin version of the HKL model, 
\begin{align}
 \tilde{\cal{H}}_{\rm HKL} & =  \sum_{\bd{k}}  \left\{ - h_{\bd{k}}  
 S^z_{ \bd{k}} 
 + \frac{U}{2}  \left[  ( S^z_{\bd{k}} )^2  - 1 \right] \right\}
 \nonumber
 \\
 & +  \frac{1}{2 \cal{V}} \sum_{\bd{k} \bd{k}^\prime } 
 f_{\bd{k} \bd{k}^\prime } S^z_{\bd{k} } S^z_{\bd{k}^\prime }  .
 \label{eq:HKLpseudo}
 \end{align}
This is a generalized Ising model where the composite Ising spins $S^z_{\bd{k}}
 = S^{z}_{ \bd{k} \uparrow} + S^z_{\bd{k} \downarrow} $ located
at the sites of the reciprocal lattice are exposed to an inhomogeneous magnetic field $h_{\bd{k}}$ and are coupled by  Landau interactions. 
In the special case where the 
interaction has zero range in reciprocal space,
$f_{\bd{k} \bd{k}^\prime} = \delta U   {\cal{V}} \delta_{\bd{k}  \bd{k}^\prime}$,
it can be absorbed into a redefinition of $U \rightarrow U + \delta U$.
Here we are interested in the case where the interaction has a finite range $ 1 / \lambda$ in reciprocal space which is however small compared with the inverse lattice spacing. 
A convenient parametrization of such an interaction in $d$ dimensions is
 \begin{align}
  f_{\bd{k} \bd{k}^\prime} & =   ( 2 \pi )^d f_0 G ( \bd{k} -
   \bd{k}^\prime ),
  \end{align}
where the function $G ( \bd{k} -
   \bd{k}^\prime )$ is the Green function of the Helmholtz equation in momentum space,
  \begin{equation}
 ( - \mathbf{\nabla}_{\bd{k}^\prime}^2   +\lambda^2 )  G (  \bd{k} - \bd{k}^\prime ) 
 =    \delta ( \bd{k} - \bd{k}^\prime ),
 \label{eq:GHelm}
 \end{equation}
with boundary condition $G ( \bd{k} - \bd{k}^\prime ) \rightarrow 0$ for 
$ | \bd{k} - \bd{k}^\prime |  \rightarrow \infty$.
Note that for $\cal{V} \rightarrow \infty$ the matrix elements of the 
inverse interaction are
\begin{equation}
  [\hat{f}^{-1}]_{\bd{k} \bd{k}^\prime} =   \frac{( 2 \pi )^d}{f_0} ( - \mathbf{\nabla}_{\bd{k}}^2 + \lambda^2 )
  \delta  ( \bd{k} - \bd{k}^\prime ),
  \label{eq:gmatrix}
  \end{equation}
where we have used the continuum normalization
 \begin{equation}
 \frac{1}{\cal{V}} \sum_{\bd{k}_1} [\hat{f}^{-1}]_{\bd{k} \bd{k}_1} f_{\bd{k}_1 \bd{k}^\prime} = {\cal{V}} \delta_{\bd{k} \bd{k}^\prime}.
 \end{equation}
In three dimensions the solution of Eq.~(\ref{eq:GHelm}) can be obtained from the usual screened Coulomb potential by substituting real space by momentum space,
 \begin{equation}
 G ( \bd{k} - \bd{k}^\prime ) = \frac{ e^{ - \lambda | \bd{k} - \bd{k}^\prime | }}{4 \pi | \bd{k} - \bd{k}^\prime | }.
 \end{equation}
In two dimensions the solution of Eq.~(\ref{eq:GHelm}) is
 \begin{equation}
 G ( \bd{k} - \bd{k}^\prime ) =   \frac{ K_0 (  \lambda | \bd{k} - \bd{k}^\prime |  )}{2 \pi},
 \end{equation}
where $K_0 (x )$ denotes the modified Bessel function with index
$n =0$, which decays exponentially for large $x$; i.e.
 $K_0 ( x ) \sim \sqrt{ \pi / ( 2 x ) } e^{- x }$ for $x \rightarrow
 \infty$, while $K_0 ( x ) \sim - \ln x$ for $x \rightarrow 0$. 
Finally, in one dimension one easily verifies
 \begin{equation}
 G ( \bd{k} - \bd{k}^\prime ) = \frac{e^{ - \lambda | \bd{k} - \bd{k}^\prime | }}{ 2 \lambda }.
 \end{equation}
In arbitrary  dimensions  $G (  \bd{k} - \bd{k}^\prime )$ is 
positive and vanishes exponentially for $ | \bd{k} - \bd{k}^\prime | \gtrsim 1/ \lambda$.

\begin{figure}[tb]
 \begin{center}
  \centering
 \includegraphics[width=0.5\textwidth]{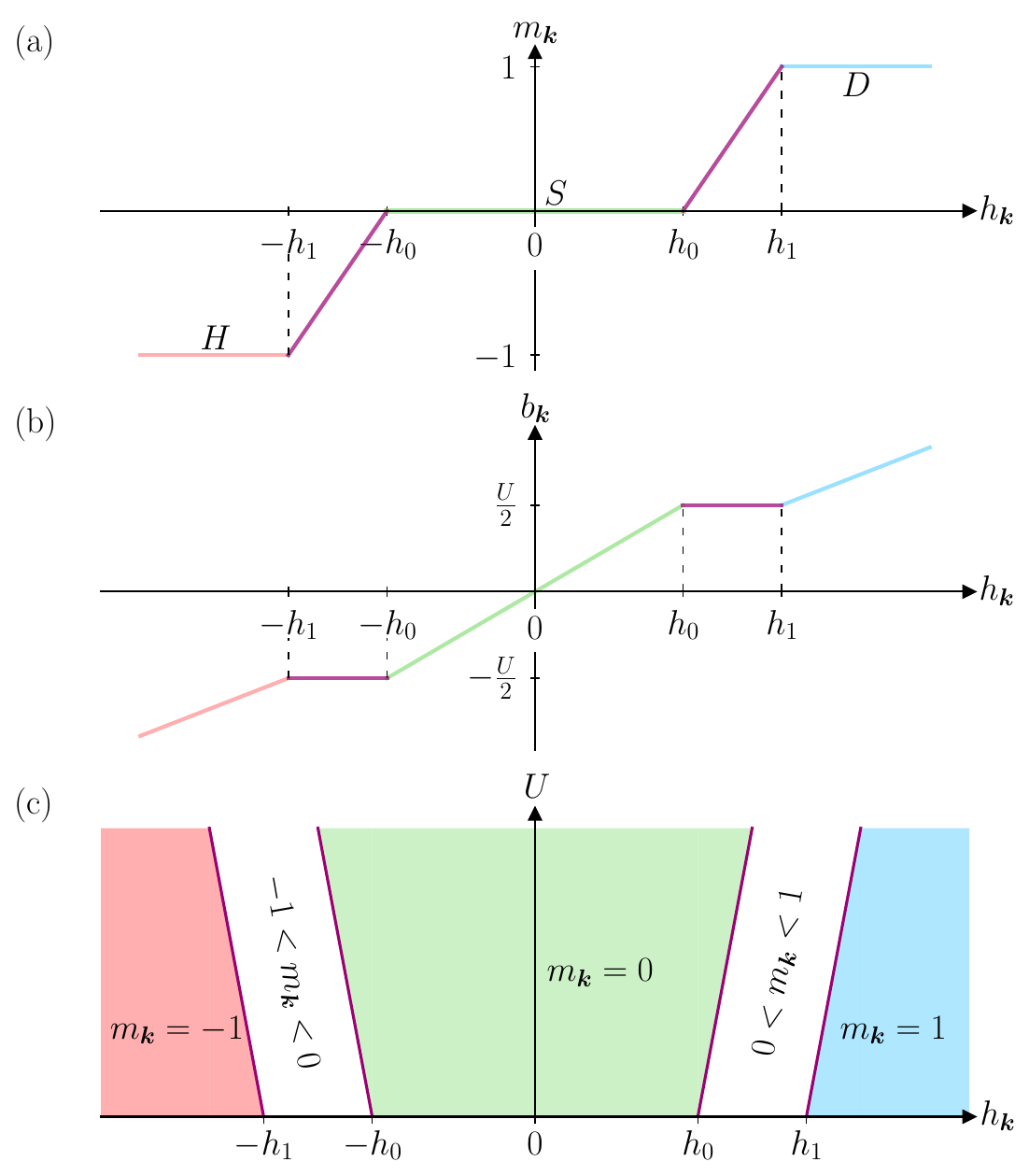}
   \end{center}
  \vspace{-4mm}
  \caption{%
(a) Local magnetic moment $m_{\bd{k}} = \langle S^z_{\bd{k}} \rangle$
of the pseudospin HKL model (\ref{eq:HKLpseudo}) as function of the
pseudomagnetic field $h_{\bd{k}}$.
(b) Effective magnetic field $b_{\bd{k}} = h_{\bd{k}}
+ \psi_{\bd{k}}$ as a function of $h_{\bd{k}}$. (c) Phase diagram of the HKL model for positive $U$ in the plane spanned by $h_{\bd{k}}$ and $U$. In the white stripes
the effective magnetic field $b_{\bd{k}}$ is frozen at  $\pm U /2$  while the magnetization increases linearly from zero to the saturation value~$\pm 1$.
}
\label{fig:mag_condens}
\end{figure}

Because the  generalized Ising Hamiltonian (\ref{eq:HKLpseudo}) is not exactly solvable, 
we apply a self-consistent mean-field decoupling, which amounts to the
replacement
 \begin{align}
\frac{1}{2 \cal{V}} \sum_{\bd{k} \bd{k}^\prime } 
 f_{\bd{k} \bd{k}^\prime } S^z_{\bd{k} } S^z_{\bd{k}^\prime } &  \rightarrow
 \frac{1}{ \cal{V}} \sum_{\bd{k} \bd{k}^\prime } 
 f_{\bd{k} \bd{k}^\prime } S^z_{\bd{k} } \langle S^z_{\bd{k}^\prime } \rangle 
 \nonumber
 \\ & -
 \frac{1}{2 \cal{V}} \sum_{\bd{k} \bd{k}^\prime } 
 f_{\bd{k} \bd{k}^\prime } \langle S^z_{\bd{k} } \rangle  \langle S^z_{\bd{k}^\prime } \rangle .
 \end{align}
At finite temperature $T = 1/ \beta$  the self-consistent magnetization
$m_{\bd{k}} = \langle S^z_{\bd{k}} \rangle$ then satisfies
\begin{align}
 m_{\bd{k}} & = 
 \frac{ \sinh ( \beta b_{\bd{k}} ) }{ \cosh ( \beta b_{\bd{k}} ) + e^{ \beta U /2} },
 \label{eq:Poissonmag}
 \end{align}
where we have introduced the self-consistent effective  magnetic field
 \begin{equation}
 b_{\bd{k}} = h_{\bd{k}} - \frac{1}{\cal{V}} \sum_{\bd{k}^\prime} f_{\bd{k} \bd{k}^{\prime} } m_{\bd{k}^\prime}.
 \label{eq:bdef}
 \end{equation} 
In the zero-temperature limit $\beta \rightarrow \infty$ the self-consistency equation 
(\ref{eq:Poissonmag})
for the pseudomagnetization $m_{\bd{k}}$ reduces to
 \begin{align}
  m_{\bd{k}}  & = 
   \Theta ( | b_{\bd{k}} | - U/2 ) {\rm sign } b_{\bd{k}}
  \nonumber
  \\
  &  = \left\{ 
 \begin{array}{cc}  
  0  & \mbox{if $| b_{\bd{k}} | < U/2$},\\
 {\rm sign } b_{\bd{k}} & \mbox{if $ | b_{\bd{k}} | > U/2$}.
\end{array} \right.
 \label{eq:mselfcon}
  \end{align}
Eqs.~(\ref{eq:bdef}) and (\ref{eq:mselfcon}) form a system of two coupled equations for the two unknowns $m_{\bd{k}}$ and $b_{\bd{k}}$.  
To construct an explicit solution, we 
introduce the auxiliary potential 
\begin{equation}
 \psi_{\bd{k}} =  b_{\bd{k}} - h_{\bd{k}} = - \frac{1}{\cal{V}} \sum_{\bd{k}^\prime} f_{\bd{k} \bd{k}^\prime} m_{\bd{k}^\prime}.
 \end{equation}
Using the fact that for our special interaction the inverse of the integral kernel $f_{\bd{k} \bd{k}^\prime }$ is given by Eq.~(\ref{eq:gmatrix}),
we find  that our auxiliary potential satisfies 
the following non-linear partial differential equation,
 \begin{align}
 & ( - \mathbf{\nabla}_{\bd{k}}^2   +\lambda^2 ) \psi_{\bd{k}} =
 - f_0  m_{\bd{k}}
 \nonumber
 \\
 & = - f_0
 \Theta ( | h_{\bd{k}} + \psi_{\bd{k}} | - U/2 )  {\rm sgn} ( h_{\bd{k}} + \psi_{\bd{k}} ) .
 \label{eq:psidif}
 \end{align}
Note that this can be viewed as a magnetic Poisson equation in momentum space.
Inspired by the fermion condensation scenario proposed by  Khodel and Shaginyan \cite{Khodel90}, we look for a solution with
$|b_{\bd{k}}| = U/2$ in some finite sector of the Brillouin zone. 
For $b_{\bd{k}} > 0$ we therefore impose the following boundary conditions
 \begin{subequations}
  \label{eq:boundarya}
 \begin{align}
 b_{\bd{k}} & = \frac{U}{2} , \; \; \; \;  \mbox{if $ \bd{k} \in {\cal{B}}^+_{\ast}$},
 \label{eq:boundarya1}
 \\
 m_{\bd{k}} & = 0 , \; \; \; \; \; \;  \mbox{if $ \bd{k} \in {\cal{B}}^+_{0}$},
 \\
 m_{\bd{k}} & = 1 , \;  \; \; \; \; \; \mbox{if $ \bd{k} \in {\cal{B}}^+_{1}$}.
 \end{align}
 \end{subequations}
while for 
$b_{\bd{k}} < 0$ the boundary conditions are
 \begin{subequations}
 \begin{align}
  b_{\bd{k}} & = - \frac{U}{2} , \; \; \; \mbox{if $ \bd{k} \in {\cal{B}}^-_\ast$},
 \\
 m_{\bd{k}} & = 0 , \; \; \; \; \; \; \; \, \mbox{if $ \bd{k} \in {\cal{B}}^-_{0}$},
 \\
 m_{\bd{k}} & = -1 , \; \; \; \; \; \mbox{if $ \bd{k} \in {\cal{B}}^-_{1}$}.
 \end{align}
 \end{subequations}
Here ${\cal{B}}^{\pm}_{\ast}$, ${\cal{B}}^{\pm}_{0}$ and ${\cal{B}}^{\pm}_{1}$
are certain sectors of the Brillouin zone such that
${\cal{B}}^+_{\ast} \cup    {\cal{B}}^+_{0} \cup {\cal{B}}^+_{1}$ is the regime where
$b_{\bd{k}} > 0$ and
${\cal{B}}^-_{\ast} \cup    {\cal{B}}^-_{0} \cup {\cal{B}}^-_{1}$ 
is the regime where
$b_{\bd{k}} < 0$.
Consider first the case $b_{\bd{k}} > 0$. The boundary condition
(\ref{eq:boundarya1}) then implies 
 \begin{equation}
  \psi_{\bd{k}} =  \frac{U}{2} - h_{\bd{k}} , \; \; \; \bd{k} \in {\cal{B}}^+_\ast.
  \end{equation}
Substituting this into Eq.~(\ref{eq:psidif}) we conclude that in this regime
 \begin{equation}
  m_{\bd{k}} = \frac{   ( - \mathbf{\nabla}_{\bd{k}}^2   +\lambda^2 ) 
  \left(
  h_{\bd{k}} - \frac{U}{2} \right)}{f_0}
  , \; \; \; \bd{k} \in {\cal{B}}^+_\ast.
  \label{eq:kBast}
  \end{equation}
The surface $ \partial {\cal{B}}^+_{0}$   of the regime ${\cal{B}}^+_{0}$ is then determined by
 \begin{equation}
 0 = \frac{   ( - \mathbf{\nabla}_{\bd{k}}^2   +\lambda^2 ) 
  \left(
  h_{\bd{k}} - \frac{U}{2} \right)}{f_0}
  , \; \; \; \bd{k} \in \partial {\cal{B}}^+_{0}.
  \label{eq:kB0}
  \end{equation}
while on the  surface of
$ \partial {\cal{B}}^+_{1}$   of the regime ${\cal{B}}^+_{1}$ we have
 \begin{equation}
 1 = \frac{   ( - \mathbf{\nabla}_{\bd{k}}^2   +\lambda^2 ) 
  \left(
  h_{\bd{k}} - \frac{U}{2} \right)}{f_0}
  , \; \; \; \bd{k} \in \partial {\cal{B}}^+_{1}.
  \label{eq:kB1}
  \end{equation}
On the other hand, for $b_{\bd{k}} < 0$ the auxiliary potential in the transition regime ${\cal{B}}^-_{\ast}$ is given by
\begin{equation}
  \psi_{\bd{k}} =  - \frac{U}{2} - h_{\bd{k}} , \; \; \; \bd{k} \in {\cal{B}}^-_\ast,
  \end{equation}
which implies
\begin{equation}
  m_{\bd{k}} = - \frac{   ( - \mathbf{\nabla}_{\bd{k}}^2   +\lambda^2 ) 
  \left(
   h_{\bd{k}} + \frac{U}{2} \right)}{f_0}
  , \; \; \; \bd{k} \in {\cal{B}}^-_\ast.
  \label{eq:kBastminus}
  \end{equation}
The surface $\partial {\cal{B}}^-_0$ of
the regime ${\cal{B}}^-_{{0}}$ is therefore given by
\begin{equation}
 0 = \frac{   ( - \mathbf{\nabla}_{\bd{k}}^2   +\lambda^2 ) 
  \left(
  h_{\bd{k}} + \frac{U}{2} \right)}{f_0}
  , \; \; \; \bd{k} \in \partial {\cal{B}}^-_{0}.
  \label{eq:kB0minus}
  \end{equation}
while on the surface $\partial {\cal{B}}^-_1$ of
${\cal{B}}^-_1$ we have
\begin{equation}
 -1 = \frac{   ( - \mathbf{\nabla}_{\bd{k}}^2   +\lambda^2 ) 
  \left(
  h_{\bd{k}} + \frac{U}{2} \right)}{f_0}
  , \; \; \; \bd{k} \in \partial {\cal{B}}^-_{1}.
  \label{eq:kB1minus}
  \end{equation}
For the nearest-neighbor tight-binding dispersion $\epsilon_{\bd{k}}$
given in Eq.~(\ref{eq:epsdef})
we obtain
 \begin{align}
 - \mathbf{\nabla}_{\bd{k}}^2 h_{\bd{k}} & = 
 \mathbf{\nabla}_{\bd{k}}^2 \epsilon_{\bd{k}}  = - a^2 
 \epsilon_{\bd{k}} = - a^2 ( \mu - h_{\bd{k}} ),
 \end{align}
so that the condition (\ref{eq:kB0}) reduces to
 \begin{equation}
 h_{\bd{k}} = \frac{U}{2} + \left( \mu - \frac{U}{2} \right)
  \frac{a^2}{ a^2 + \lambda^2} \equiv h_0, 
   \; \; \; \bd{k} \in \partial {\cal{B}}^+_{0},
   \end{equation}
which is equivalent to
 \begin{equation}
 \epsilon_{\bd{k}} = \left( \mu - \frac{U}{2} \right)
 \frac{ \lambda^2}{a^2 + \lambda^2 },
 \; \; \; \bd{k} \in \partial {\cal{B}}^+_{0}.
 \end{equation}
This defines a surface  $ \partial {\cal{B}}^+_0$ in the Brillouin zone that is one of the boundaries
of the regime ${\cal{B}}^+_{\ast}$.
On the other hand, the condition (\ref{eq:kB1}) for the surface 
$\partial {\cal{B}}^+_1$ 
of the regime ${\cal{B}}^+_1$ can be written as
 \begin{equation}
 h_{\bd{k}} = h_0 + \frac{f_0}{a^2 + \lambda^2} \equiv h_1, 
   \; \; \; \bd{k} \in \partial {\cal{B}}^+_{1}.
   \end{equation}
Finally, from Eq.~(\ref{eq:kBast}) we conclude that
in the transition  regime
$\bd{k} \in {\cal{B}}^+_\ast$ the magnetization is simply given by
 \begin{equation}
 m_{\bd{k}} = \frac{ h_{\bd{k}} - h_0}{ h_1 - h_0 },
 \; \; \; h_1 \geq h_{\bd{k}} \geq h_0 .
 \end{equation}
We illustrate the resulting magnetic equation of state in Fig.~\ref{fig:mag_condens} (a) and the corresponding effective magnetic field $b_{\bd{k}}$  in Fig.~\ref{fig:mag_condens} (b).
The constant value of $b_{\bd{k}}$ in the transition regions 
${\cal{B}}_{\ast}^{\pm}$ means in the original fermionic picture that the
effective fermion dispersion is completely flat, which is the fermion condensation scenario
proposed by  Khodel and Shaginyan \cite{Khodel90}.
In the pseudospin picture of the HKL model, fermion condensation translates into
smoothing of the magnetic domain walls due to momentum mixing. The resulting phase diagram in the plane spanned by the pseudomagnetic field $h_{\bd{k}}$ and the interaction $U$ is shown
in Fig.~\ref{fig:mag_condens} (c).

\section{Summary and conclusions}
\label{sec:summary}

In this work we have considered  a generalization of the HK model with
momentum mixing due to Landau interactions (HKL model). 
Motivated by Andersons pseudospin formulation of BCS theory \cite{Anderson58}
we have  shown that the HKL model can be mapped 
onto a generalized Ising model in reciprocal space where two Ising spins are attached to each site of the reciprocal lattice.  In the pseudospin picture, the energy dispersion of the original HK model 
corresponds to an inhomogenuous magnetic field in momentum space, while the Fermi surfaces are represented by  magnetic domain walls. 
A self-consistent mean-field analysis of the pseudospin HKL model shows that momentum mixing destroys the discontinuities of domain walls; in the original HKL model
this corresponds fermion condensation and partially flat bands.  
Guided by the  pseudospin picture, we have also proposed an exactly solvable generalization of the HK model which does not exhibit any ground state degenercies.

The question whether 
self-consistent mean-field theory is sufficient to establish the phenomenon of fermion condensation is beyond the scope of this work.   Recall that shortly after  Khodel and Shaginyans original proposal~\cite{Khodel90}
of the fermion condensation scenario,  
Nozi\`{e}res~\cite{Nozieres92} has argued that fermion condensation
is an artifact of the mean-field approximation, which does not survive when correlation effects such as the damping of quasi-particles is properly taken into account. 
On the other hand, the proponents of the fermion condensation theory argue that it resolves many puzzling features observed in strange metal and heavy fermion 
materials \cite{Khodel1994, Khodel2005, Shaginyan2010, Shaginyan2013, Shaginyan2019, Volovik2019}. 
Moreover, recent non-perturbative calculations using a strong-coupling functional renormalization group  approach suggest that flat but significantly broadened bands can indeed exist in extremely  strongly correlated electronic systems \cite{Arnold25a,Arnold26b}.
To further clarify this issue, one should investigate the pseudospin HKL model
(\ref{eq:HKLpseudo}) beyond mean-field theory, for example by means of renormalization group methods.

Finally, let us point out that a general map between interacting  fermions and generalized Ising models has recently been constructed by Wetterich \cite{Wetterich17}. Possibly this map can  be useful to investigate the effect of fluctuations on the fermion condensation scenario for general interactions.

\section*{ACKNOWLEDGEMENTS}
	
This work was financially supported by the
Deutsche Forschungsgemeinschaft (DFG, German Research Foundation) through Project No. 431190042.

\begin{appendix}

\section*{APPENDIX: Momentum distribution of the HK model and magnetization of the pseudospin HK model}

\setcounter{equation}{0}
\renewcommand{\theequation}{A\arabic{equation}}
 
Using the fact that in the occupation number basis (\ref{eq:basisoccupation})
the Hamiltonian (\ref{eq:HK}) of the HK model is diagonal with momentum blocks given by Eq.~(\ref{eq:HKmatrix1}), the grand canonical partition function for fixed inverse temperature $\beta = 1/T$ and chemical potential $\mu$ can be immediately written down,
 \begin{align}
 {\cal{Z}}_{\rm HK} & = \prod_{\bd{k}} {\rm Tr} e^{ - \beta [ {\cal{H}}_{\bd{k}}  - \mu ( n_{\bd{k} \uparrow} + n_{\bd{k} \downarrow} ) ]}
 \nonumber
 \\
  & = \prod_{\bd{k}} \left[ 1 + 2 e^{ - \beta   \xi_{\bd{k}}} + e^{ - \beta ( 2 \xi_{\bd{k}}  + U )}
  \right],
 \end{align}
where $\xi_{\bd{k}} = \epsilon_{\bd{k}} - \mu $.
The corresponding grand canonical potential is
 \begin{align}
 \Omega_{\rm HK} & = - T \sum_{\bd{k}} \ln \left[
 1 + 2 e^{ - \beta \xi_{\bd{k}} } + e^{ - \beta ( 2 \xi_{\bd{k}} + U ) } \right].
 \end{align}
Taking a derivative with respect to $\mu$, we obtain the dimensionless density
 \begin{align}
 \rho = - \frac{1}{\cal{N}} \frac{ \partial \Omega_{\rm HK}}{\partial \mu } =
 \frac{2}{\cal{N}} \sum_{\bd{k}} \frac{  e^{ - \beta \xi_{\bd{k}} }
 [ 1 + e^{ - \beta ( \xi_{\bd{k}} + U ) } ]}{
 1 +  e^{ - \beta \xi_{\bd{k}} } [ 2 + e^{ - \beta ( \xi_{\bd{k}} + U ) } ]},
 \end{align}
which can be written as a sum over  the average occupation numbers $\bar{n}_{\bd{k}}$ 
of the sites $\bd{k}$ of the reciprocal lattice,
 \begin{equation}
\rho = \frac{1}{\cal{N}} \sum_{\bd{k}}  \bar{n}_{\bd{k}} ,
 \end{equation}
with
 \begin{equation}
 \bar{n}_{\bd{k}} = 
 \frac{  2 e^{ - \beta \xi_{\bd{k}} }
 [ 1 + e^{ - \beta ( \xi_{\bd{k}} + U ) } ]}{
 1 +  e^{ - \beta \xi_{\bd{k}} } [ 2 + e^{ - \beta ( \xi_{\bd{k}} + U ) } ]}.
  \end{equation}
In the limit $T \rightarrow 0 $ this reduces to
 \begin{subequations}
 \begin{align}
 \bar{n}_{\bd{k}} & = \left\{
  \begin{array}{cc}
 2 \Theta ( \mu - \epsilon_{\bd{k}} - U / 2 ), & 
 \mbox{for $ U < 0$}, \\
   \Theta ( \mu - \epsilon_{\bd{k}} )
 + \Theta ( \mu - \epsilon_{\bd{k}} - U ) , & 
 \mbox{for $ U > 0 $,} \end{array} \right. 
 \label{eq:nkzero}
 \end{align}
 \end{subequations}
see Eq.~(\ref{eq:barn}) of the main text. 
Hence, in the zero temperature limit
the relation between density and chemical potential of the HK model
reduces to
 \begin{subequations}
\begin{align}
 \rho & = \frac{2}{\cal{N}} \sum_{\bd{k}} \Theta ( \mu - \epsilon_{\bd{k}} - U/2 ) , \; \; \; 
 \mbox{for $ U < 0 $},
\end{align}
and 
\begin{align}
 \rho & = \frac{1}{\cal{N}} \sum_{\bd{k}} [ \Theta ( \mu - \epsilon_{\bd{k}} )
 + \Theta ( \mu - \epsilon_{\bd{k}} - U ) ] , \; \; \; 
 \mbox{for $ U > 0 $}.
 \end{align}
 \end{subequations}

Next, consider the modified HK model in the pseudospin picture 
with Hamiltonian $\tilde{\cal{H}}_{\rm HK}$ given 
in Eq.~(\ref{eq:HKpseudo}). The partition function is now
 \begin{align}
 \tilde{\cal{Z}}_{\rm HK} & = \prod_{\bd{k}} {\rm Tr} e^{ - \beta \tilde{\cal{H}}_{\bd{k}} }
  \nonumber
  \\
  & = \prod_{\bd{k}} \left[ 2 \cosh ( \beta h_{\bd{k}} ) + 2 e^{ \beta U/2 } \right],
\end{align}
so that the grand canonical potential is
 \begin{equation}
 \tilde{\Omega}_{\rm HK} = - T \sum_{\bd{k}} \ln \left[ 2 \cosh ( \beta h_{\bd{k}} ) + 2 e^{ \beta U/2 } \right].
 \end{equation}
The average magnetic moment at momentum $\bd{k}$ is therefore
 \begin{equation}
 m_{\bd{k}} = - \frac{1}{\cal{N}}  \frac{ \partial \tilde{\Omega}_{\rm HK}}{\partial h_{\bd{k}} }
=  \frac{ \sinh ( \beta h_{\bd{k}} ) }{ \cosh ( \beta h_{\bd{k}} ) + e^{ \beta U /2} }.
\end{equation}
For $\beta \rightarrow \infty$ this reduces to  Eq.~(\ref{eq:mkres}) of the main text.

\end{appendix}

\end{document}